\documentclass[preprint]{aastex}

\tighten

\newcommand{\asec}{\hbox to 1pt{}\rlap{$^{\prime\prime}$}.\hbox to 2pt{}}
\newcommand{\amin}{\hbox to 1pt{}\rlap{$^{\prime}$}.\hbox to 2pt{}}
\newcommand{\Ser}{S\' ersic\ }

\shortauthors{Lauer et al.}
\shorttitle{Bimodal Central Brightness Profiles}

\begin{document}

\title{The Centers of Early-Type Galaxies with HST.
VI. Bimodal Central Surface Brightness Profiles.
\footnote{Based on observations made with the NASA/ESA 
{\it Hubble Space Telescope}, obtained at the Space Telescope Science Institute,
which is operated by the Association of Universities for
Research in Astronomy, Inc., under NASA contract NAS 5-26555. These
observations are associated with GO and GTO proposals
\# 5236, 5446, 5454, 5512, 5943, 5990, 5999, 6099, 6386, 6554, 6587, 6633,
7468, 8683, and 9107.}}

\author{Tod R. Lauer}
\affil{National Optical Astronomy Observatory\footnote{The National Optical
Astronomy Observatory is operated by AURA, Inc., under cooperative agreement
with the National Science Foundation.},       
P.O. Box 26732, Tucson, AZ 85726}

\author{Karl Gebhardt}
\affil{Department of Astronomy, University of Texas, Austin, Texas 78712}

\author{S. M. Faber}
\affil{UCO/Lick Observatory, Board of Studies in Astronomy and
Astrophysics, University of California, Santa Cruz, California 95064}

\author{Douglas Richstone}
\affil{Department of Astronomy, University of Michigan, Ann Arbor, MI 48109}

\author{Scott Tremaine}
\affil{Princeton University Observatory, Peyton Hall, Princeton, NJ 08544}

\author{John Kormendy}
\affil{Department of Astronomy, University of Texas, Austin, Texas 78712}

\author{M. C. Aller}
\affil{Department of Astronomy, University of Michigan, Ann Arbor, MI 48109}

\author{Ralf Bender}
\affil{Universit\"{a}ts-Sternwarte, Scheinerstra\ss e 1, M\"{u}nchen 81679,
Germany}

\author{Alan Dressler}
\affil{The Observatories of the Carnegie Institution of Washington,
813 Santa Barbara St., Pasadena, CA 91101}

\author{Alexei V. Filippenko}
\affil{Department of Astronomy, University of California, Berkeley,
CA 94720-3411}

\author{Richard Green}
\affil{LBT Observatory, University of Arizona, Tucson, AZ 85721}

\author{Luis C. Ho}
\affil{The Observatories of the Carnegie Institution of Washington,
813 Santa Barbara St., Pasadena, CA 91101}

\vfill


\begin{abstract}

We combine the results from several {\it HST} investigations of
the central structure of early-type galaxies to generate a large sample
of parameterized surface photometry.
The studies selected for inclusion were those that used the ``Nuker law"
to characterize the inner light distributions of the galaxies.
The sample comprises
WFPC1 and WFPC2 $V$ band observations published earlier by our group,
$R$ band WFPC2 photometry of Rest et al., NICMOS $H$ band photometry
by Ravindranath et al.\ and Quillen et al., and the brightest cluster
galaxy WFPC2 $I$ band
photometry of Laine et al.  All parameters are transformed
to the $V$ band and a common distance scale.
The distribution of the logarithmic slopes of the
central brightness profiles strongly affirms that the central structure
of elliptical galaxies with $M_V<-19$ is bimodal,
based on both parametric and non-parametric analysis.
At the {\it HST} resolution limit,
most galaxies are either power-law systems, which have steep cusps
in surface brightness, or core systems, which have
shallow cusps interior to a steeper envelope brightness distribution.
A rapid transition between the two forms occurs over the luminosity range
$-22<M_V<-20,$ with cores dominating at the highest luminosities,
and power-laws at the lowest.  There are a few ``intermediate'' systems
that have both cusp slopes and total luminosities that fall within the
core/power-law transition, but they are rare and do not fill in the
overall bimodal distribution of cusp slopes.  These results are inconsistent
with the Ferrarese et al. Virgo Cluster Survey (VCS) analysis, which
did not find a bimodal distribution of cusp slopes.
While the VCS ACS/WFC images have lower angular resolution than the
WFPC2/PC F555W images used for much of the sample, the basic galaxy profiles
measured with either camera agree well after deconvolution.
However, using galaxies common to the present and VCS samples,
we demonstrate that the VCS models of the cusps are either a poor match
to the observations or consist of forms fitted to the galaxy envelopes and
extrapolated inward to the {\it HST} resolution limit.

\end{abstract}


\section{Introduction} 

Long before massive black holes were accepted to be present in the
centers of galaxies, \citet{bbr} argued that a binary black hole
created in the merger of two galaxies would eject stars from the center of
the newly created system as the binary slowly hardened.
N-body simulations support this hypothesis \citep{ebi, mak97, mnm}.
We now believe that nearly every elliptical galaxy
or spiral bulge has a black hole at its center \citep{mag}. 
We also believe that the most massive elliptical galaxies were formed
by merging pre-existing galaxies.  It is possible that their
central structure still bears witness to such events.
The ``cores'' seen at the centers of the most luminous elliptical galaxies
may indeed be the signatures of gravitational stirring and heating (``core
scouring'') by binary black holes \citep{f97}.

Cores were initially seen in ground-based observations of
nearby luminous elliptical galaxies
as a central region of nearly constant surface brightness \citep{l85, k85}.
The cores were seen to be
non-isothermal, but their true form as $r\rightarrow0$ was unknown.
{\it HST} images, however, later showed that nearly
all galaxies have singular starlight distributions in the sense that
surface brightness diverges as $\Sigma(r)\sim r^{-\gamma}$
\citep{l91, l92a, l92b, crane, k94, f94, l95}.
Typically, in low luminosity
early-type galaxies, $\gamma$ decreases only slowly as the center is approached
and a steep $\gamma>0.5$ cusp continues into the {\it HST} resolution limit;
\citet{l95} classified these systems as ``power-law'' galaxies.  In more
luminous galaxies, however, the steep envelope profile transitions
to a shallow inner cusp with $\gamma<0.3$ at a ``break radius,'' $r_b.$
This behavior was seen in galaxies that had cores
from ground-based observations; the break radius roughly
corresponded to what had been measured as the core radius.
We thus chose to continue to
call these ``core galaxies,'' even though the 
shallow cusps in projected brightness in most\footnote{\citet{l02} and
\citet{l05} identified a number of galaxies in which the brightness
profiles actually decreased as $r\rightarrow0.$  These systems thus have
$\gamma<0$ at small radii and do not have cusps in space density.}
of these systems imply steep and
singular cusps in luminosity density \citep{l95}.

A strong justification of this classification schema was that
the distribution of $\gamma'$ over a sample of early-type galaxies
was seen to be strongly bimodal \citep{f97},
where $\gamma'$ is the local cusp slope at the {\it HST} resolution limit.
Initially no galaxies were seen to have $0.3<\gamma'<0.5.$
It was thus sensible to apply the separate core and power-law
classifications to the two distinct and cleanly separated peaks
in the $\gamma'$ distribution.
The bimodal distribution was initially inferred from fits
to the surface photometry using the ``Nuker law'' (\citealt{l95};
see equation \ref{eqn:nuker}), which
has the asymptotic form $\Sigma(r)\sim r^{-\gamma},$ as $r\rightarrow0,$
however, \citet{g96} demonstrated that non-parametric treatment
of the surface photometry recovered an identical bimodal distribution,
and further that bimodality
was a feature of the luminosity {\it density} distributions of
early type galaxies, not just the surface brightness distributions.

The core and power-law classifications are fundamental because they correlate
with global physical properties that are diagnostic of galaxy formation.
Core galaxies are on average more luminous than power-law galaxies,
with power-laws dominating below $M_V\sim-21$ and cores greatly
dominating at higher luminosities \citep{f97}.
Core galaxies generally have boxy isophotes, while power-law galaxies
are disky \citep{n91, k96, f97}.
Core galaxies rotate slowly, while power-law galaxies rotate more rapidly
\citep{f97}; similarly, boxy galaxies include slow rotators while 
disky galaxies rotate rapidly \citep{b88, b89, b94, k96}.
Core-boxy galaxies show minor-axis rotation indicative 
of triaxiality, while coreless-disky galaxies do not \citep{k96}.
And core galaxies are systematically rounder than power-law galaxies
\citep{jaffe, f94, ryden2, l05}, just as
bright (boxy) galaxies are systematically rounder than faint (disky)
galaxies \citep{tm}.
Significantly, core and power-law galaxies of identical luminosity
can still be separated by all of these secondary characteristics \citep{f97}.
Thus central and global properties together point to
differences in the formation of core and power-law ellipticals.
\citet{k96} propose a revision of the Hubble sequence for
elliptical galaxies that recognizes this
physical dichotomy, using isophote shape as a
convenient surrogate for the above physical properties.  More recent
work continues to show that power-law and core galaxies have physically
distinct properties.  Cores \citep{cap, bal},
like boxy isophotes \citep{b89}, correlate with radio-loud active nuclei.
And cores \citep{pell}, like
boxy isophotes \citep{b89}, correlate with strong x-ray emission.

The properties of the power-law versus core galaxies
suggest that core systems resulted from the mergers
of less luminous power-law systems.
Simulations by \citet{mnm}, for example, showed that the merger of two power-law
galaxies, each with a central massive black hole, would produce a core galaxy.
The bimodal cusp distribution would thus be one key to understanding
the central structure of elliptical galaxies over a history of mergers.

As larger samples of elliptical galaxies were observed with {\it HST},
the bimodal slope distribution remained robust \citep{quil, rest, rav, l05}.
While all of these studies identified a small number of ``intermediate''
galaxies, which have limiting cusp slopes with $0.3<\gamma'<0.5,$
they are not common enough to fill the ``valley'' between the
power-law and core cusp-slope distributions.  This is a critical point
that has often been misunderstood in the literature.  The sample
of galaxies available to \citet{f97} had no systems
with $0.3<\gamma'<0.5,$ thus the distributions of slopes was
not only bimodal, but disjoint.
The subsequent discovery of a small number of intermediate galaxies shows
that the distribution of cusp slopes is truly continuous, but it has remained
bimodal in its overall character in all the studies cited above.

This picture has been challenged by the recent distribution of cusp
slopes presented by \citet{lf}, who do not find bimodality
in their sample of 100 galaxies observed by their {\it HST}
Virgo Cluster Survey (VCS), although
they still identify a separate class of core galaxies, which
they define as having brightness profiles that fall below
\citet{sersic} forms in the center.
While the use of \Ser profiles is a new approach to
identifying core galaxies, the class of VCS core galaxies identified
in this way is essentially consistent with the definition of core
galaxies in earlier studies.
The distribution of slopes for the non-core galaxies in \citet{lf},
however, does not match that of the power-law galaxies presented
in all the studies cited above,
even though this set includes several galaxies previously
classified as power-laws.  The difference between
the VCS work and all earlier studies lies not in whether or not there are
two classes of galaxies, core and "non-core", but in the form of the
central structure of the non-core class.
Resolving the different pictures presented by
\citet{f97} and the VCS results is crucial to understanding
how the central structure of elliptical galaxies varies with luminosity,
and how it may change over mergers.

This paper has three goals.  We first construct a large sample
of galaxies that have had their surface brightness distributions
characterized by Nuker law fits.  These are reduced to a common photometric
and distance scale.  This sample provides the basic material for an
updated investigation into the relationships between the central
structure of galaxies and their more global properties, as was presented
in \citet{f97}; this new analysis is presented in \citet{l07}.
The second goal is to reinvestigate the distribution of cusp slopes
in early type galaxies.  We find that the distribution of slopes remains
strongly bimodal, a conclusion that we ratify with a repeat of the
non-parametric analysis presented by \citet{g96}.
The final task is to bolster this conclusion
with a comparison of the analysis of
{\it HST} imagery by \citet{l95, l05} to that
of \citet{lf}.  We show that the VCS surface photometry models do
not accurately recover the distribution of $\gamma'$
for galaxies with $M_V<-19.$

\section{Compilation of a Large Sample of Early-Type Galaxy Central
Structure\label{sec:samp}}

We define a combined sample of 219 early-type galaxies observed with
{\it HST} for which Nuker-law parameters have been derived by a variety
of studies \citep{l95, f97, quil, rest, rav, laine, l05}.
The global properties of the sample galaxies are listed in Table \ref{tab:glob}.
The Nuker law parameterizations of the central structure
of the galaxies are listed in Table \ref{tab:cen}.

The investigations cited above were generally done in different photometric
bands, and used a variety of sources for galaxy distances, luminosities,
and so on.  We have adopted the F555W or {\it V} band as the common photometric
system, which was the choice of \citet{l95}, \cite{f97}, and \citet{l05}.
We also specify the absolute galaxy luminosities in the {\it V} band.
Luminosities are based on the $V_T$ apparent magnitudes published
in the RC3 \citep{rc3}, with the exception of a few galaxies that
had to be transformed from $B_T,$ and the \citet{laine} BGC sample,
which provided total magnitudes in the $R$ band.
Morphologies are also provided by the RC3.  Many of the galaxies in the
sample are S0 systems rather than true ellipticals.  As
bulge parameters, rather than those of the whole galaxy, are relevant
to the properties of central black holes \citep{mag,g00,kg01}, we have corrected
(when possible) the total luminosities of the S0 galaxies for their
disk components.  Most of the corrections were based on the bulge-disk
decompositions of \citet{bag}; other decompositions were provided
by \citet{k83}, \citet{sd}, \citet{kent}, \citet{bor}, and \citet{bur}.  When
a galaxy appeared in multiple sources we selected the median disk correction.
The final disk corrections and sources are given in Table \ref{tab:glob}.

Calculation of absolute magnitudes is
complicated by the heterogeneous quality of distance information
across the sample.  We set the overall scale to
$H_0=70$ km s$^{-1}$ Mpc$^{-1}$ and have reworked the
distances in all references.  Our preferred source of distances is the
\citet{ton} surface brightness fluctuation (SBF) survey, but using the
group memberships in \citet{f89} and averaging the SBF distances
over the group.  As the \cite{ton} SBF scale is consistent with
$H_0=74,$ however, we scale up their SBF distances by 6\%.  For galaxies
not in the SBF catalogue but that are in the \citet{f97} sample,
we rescale the distances in that paper from its scale of $H_0=80.$
For galaxies not in either sample, we check for possible membership
in the \citet{f89} group catalogues, using the group velocity in the
cosmic microwave background (CMB) frame, if a group SBF distance is not
available.  Lastly, if no group information is available, we use the
galaxy's velocity in the CMB frame.  The distances adopted are given
in Table \ref{tab:glob}; the distance references listed correspond
to the four choices, in order of priority, that we have listed here.
Absolute luminosities were calculated using the \citet{sfd} Galactic
extinction values.

Table \ref{tab:glob} also lists the major-axis effective radii
of the galaxies.
Effective radii were determined by fitting
$r^{1/4}$-laws to ground-based surface photometry profiles
available in the published literature (Lauer et al; in preparation);
unfortunately, supporting ground-based surface photometry could
not be located for 30\%\
of the sample.  No attempt was made to fit S0 galaxies with separate bulge
and disk components, so the effective radii may be suspect when the disk
contributes significantly to the total galaxy luminosity.
Under these general guidelines, we discuss the specific details and
treatment applied to each data source for the sample.  The samples
are given in order of preference, as will be explained in $\S\ref{sec:mult}$

\subsection{The \citet{l05} WFPC2 Sample}

\citet{l05} present images and surface photometry for 77 early-type galaxies
observed with {\it HST}+WFPC2, with the galaxies centered in the high-resolution
PC1 CCD; 55 of these were obtained under programs
GO 5512, 6099, 6587, and 9107, which were carried out by our collaboration.
\citet{l05} also include an independent reduction of the galaxies observed
in GO 5454 \citep{carollo}.
There is no single criterion that characterizes this combined sample.
In general, it comprises relatively luminous nearby elliptical and
early-S0 galaxies that could be investigated spectroscopically with
{\it HST} to detect and ``weigh" nuclear black holes.  The
selection of galaxies was heavily influenced by the central-structure
parameter relationships presented by \citet{f97}. Several galaxies were
chosen to sample the luminosity range over which power-law
and core galaxies coexist, to explore the transition between the two types.
Others were selected to extend the luminosity range of the core
parameter relationships.  Overall, the sample is rich in core galaxies,
with relatively few power-law galaxies; thus it is complementary to the
\citet{rest} sample, which is rich in power-law galaxies.  The adopted
resolution limit, which is the scale at which $\gamma'$ is
evaluated, is $r_0=0\asec04$ for most galaxies, but
$r_0=0\asec02$ for some systems for which double-sampled images were available.

\subsection{The \citet{rest} WFPC2 Sample}

\cite{rest} observed 68 early-type galaxies in the F702W filter with
WFPC2 PC1.  Their sample comprises early-type galaxies within
3400 km s$^{-1},$ $M_V<-18.5,$ and Galactic latitude $>20^\circ.$
Rest et al. measured surface photometry from deconvolved images
and fitted Nuker laws to the profiles.  We adopted the parameters
in Rest et al., using $F702W-F555W=0.61$ to transform the surface photometry.
One caveat is that Rest et al. adopted $r_0=0\asec1$ as the resolution
limit of their profiles, using the cusp slope at this radius for
profile classification.  We consider this resolution limit to be
unduly conservative for deconvolved WFPC2 images, and thus
used their Nuker fits at $r_0=0\asec04$ instead to determine $\gamma'.$
As a result, all ``intermediate'' galaxies in Rest et al.
are considered here to have resolved cores, while a number of their
power-law galaxies are now classified as intermediates
(see the note column in Table \ref{tab:cen}).

\subsection{The \citet{laine} WFPC2 BCG Sample}

\citet{laine} obtained WFPC2 PC1 images of 81 brightest cluster galaxies
(BCGs) in the F814W filter.  The BCG sample was drawn from the volume-limited
BCG sample of \citet{pl}.  The photometry was derived from deconvolved images;
the Nuker law parameters are combined into the sample, using
$V-I=1.4$ to transform the F814W photometry to F555W, and $V-R=0.5$
to transform the $R$ total galaxy apparent magnitudes listed in \citet{laine}.
BCG distances were based on the CMB-frame cluster velocities
tabulated by \citet{pl}.

\subsection{The \citet{l95} and \citet{f97} WFPC1 Sample}

The core parameter relationships presented in \citet{f97} were
largely based on the \citet{l95} photometry, but supplemented
with a small set of otherwise unpublished WFPC1 GTO galaxies
observed and reduced in parallel with the \citet{l95} galaxies.
The \citet{l95} sample was observed with the F555W filter in
the high-resolution PC6 CCD of WFPC1.  The surface photometry
was measured from deconvolved images; as discussed in \citet{l05},
the profiles should be reliable into $r_0\approx0\asec1.$
\citet{byun} published Nuker-law fits for all the \citet{f97} galaxies.
However, we have re-fitted the photometry for this paper to allow for
$\gamma<0,$ and in some cases have restricted the radial range of the fits
to better isolate the break radius and inner cusp.
The new parameters are also referenced to the major-axis of the galaxies
rather than to their mean-radii as was done in \citet{byun}.

\subsection{The \citet{quil} NICMOS Sample}

\citet{quil} observed 27 early-type galaxies in the $H$ band with NICMOS2;
the sample was largely based on bright nearby elliptical galaxies.
Quillen et al.\ deconvolved their photometry and fitted Nuker laws
to the profiles.  We have transformed the Quillen et al.\ photometry
using $V-H=2.95.$  We consider the limiting resolution of
the NICMOS sample to be $r_0=0\asec1.$

\subsection{The \citet{rav} NICMOS Sample}

\citet{rav} presented $H$ band photometry of 33 early-type galaxies
based on archival NICMOS2 and NICMOS3 images, many of which were also
included in the \citet{quil} sample.  The sample selection
in part required galaxy apparent magnitude $B_T<12.5$ and a northern
declination.  Ravindranath et al. did not deconvolve their images
but chose instead to fit PSF-convolved Nuker laws directly
to the images; they also included a central point-source in their fits.
We have transformed the Ravindranath et al. photometry
to F555W using $V-H=2.95$ and $B-V=0.95$ for galaxy apparent magnitudes.
Again we use $r_0=0\asec1$ as the limiting resolution of this sample.

\subsection{Galaxies That Appear in Multiple Sources\label{sec:mult}}

There is significant overlap between the six sources
that contribute to the sample.
Given the somewhat heterogeneous nature of the
source material and reduction techniques, we have chosen to rank
the sources in a preferred order, rather than to average or otherwise blend
the parameters for any galaxy observed in more that one sample.
Our first preference is for the observations presented by \citet{l05};
as these were derived from F555W WFPC2 images, they have the best
spatial resolution of the six studies used.
Our next preferences are for the the \citet{rest} F702W profiles,
and then the \citet{laine} F814W profiles, as they have successively
lower resolution, given the redder filters.
While WFPC2 offers superior resolution to WFPC1, deconvolved WFPC1 imagery
should still be superior to NICMOS observations made in the near-IR,
so the galaxies in \citet{l95} or \citet{f97} that have no
WFPC2 imagery are then next in order of preference.  Of
the two NICMOS programs, we prefer photometry from \citet{quil} as they
fitted photometry from the deconvolved images directly, rather than
fitting PSF-convolved models, which are more vulnerable to systematic errors.
The NICMOS programs also provided surface brightness parameters
for a number of galaxies that were observed by WFPC2,
but that had centers so strongly affected by dust absorption that
accurate Nuker law parameters could not be measured in optical filters.

\section{The Bimodality of Central Structure and its Classification\label{sec:bimodal}}

\subsection{Core and Power-Law Galaxies}

As in our our previous analyses of the central structure of early-type galaxies
\citep{k94, l95, f97}, we classify the systems into two types, core
and power-law galaxies, based on the logarithmic slopes, $\gamma',$ of
their central cusps at the {\it HST} angular resolution limit, where
\begin{equation}
\gamma'(r_0)~\equiv~-{d\log I\over d\log r}\Biggr|_{r=r_0},
\end{equation}
and $r_0$ is the {\it HST} resolution scale or any other limiting radius.
In the analysis that follows, we represent $I(r)$
with the ``Nuker law'' \citep{l95} parametric form
\begin{equation}
I(r)=2^{(\beta-\gamma)/\alpha}I_b\left({r_b\over r}\right)^{\gamma}
\left[1+\left({r\over r_b}\right)^\alpha\right]^{(\gamma-\beta)/\alpha},
\label{eqn:nuker}
\end{equation}
which describes the profiles as a ``broken power law;''
it is an extension of a broken power-law form first used to
describe the brightness profile of M32 \citep{l92b}.
Note that $\gamma$ is the inner cusp slope
as $r\rightarrow0$ and is distinguished from $\gamma',$ which
is the purely local slope evaluated at the resolution limit noted above.

Core galaxies are those that exhibit a well defined
``break radius,'' $r_b,$ which marks a rapid transition (moderated
by $\alpha$) from the outer
steep envelope brightness profile (which has asymptotic slope
$\beta$) to a shallow cusp in surface brightness
that generally persists as $r\rightarrow0.$
The break radius also marks the point of maximum curvature
of the profile in logarithmic coordinates, and can thus be
readily recognized by non-parametric means in addition to the
use of the Nuker law.
In quantitative terms, core galaxies have
$\gamma'<0.3.$  The present sample contains 117 core galaxies.
 
Power-law galaxies retain a steep slope into
the resolution limit, having $\gamma'>0.5.$  These systems are
also well described by the Nuker law, but the breaks are poorly
defined with small $\alpha$, or have $\gamma$ only
slightly smaller than $\beta.$  The ``power-law'' appellation
reflects the appearance of the brightness profiles of these systems
in logarithmic coordinates.  None of them are truly constant
power-laws; they generally have modest decreases in slope as $r\rightarrow0.$
The present sample contains 89 power-law galaxies.

As \cite{l95} emphasized, we prefer to consider power-laws as
galaxies that have no core resolved, rather than presuming that the
galaxies intrinsically lack cores.
A small number of galaxies classified as power-laws from WFPC1 or NICMOS
observations, for example, did reveal cores with the improved
resolution of WFPC2 (see the notes in Table \ref{tab:cen}).
The power-law classification thus depends on context.
Since most cores measured so far have $r_b<1$ kpc, clearly the physical
resolution must be much better than this for the non-resolution of a core,
and hence the power-law classification, to be interesting.
Conversely, the power-law classification has proved to be most
intriguing for those galaxies with upper limits
on $r_b$ well below those typical to galaxies of similar luminosity.

The sample also contains 13 galaxies that have $0.3<\gamma'<0.5,$
which, following \citet{rest}, are classified as ``intermediate'' galaxies.
These systems are discussed further in $\S\ref{sec:inter}.$
 
\subsection{Tests for Bimodality}

As noted in the introduction, bimodality has been confirmed in
work conducted subsequent to \citet{f97}, with the exception of the
\citet{lf} VCS results.
With the our combined large sample,
we can now revisit the overall distribution of $\gamma'$
and retest the hypothesis that core and power-law
galaxies can be described as two distinct populations.
This is especially interesting as
work by \citet{rest}, \citet{rav}, and \citet{l05}
has identified a number of ``intermediate'' galaxies that
have $0.3<\gamma'<0.5,$ thus falling between the core and power-law classes.

\begin{figure}
\centering
\plotone{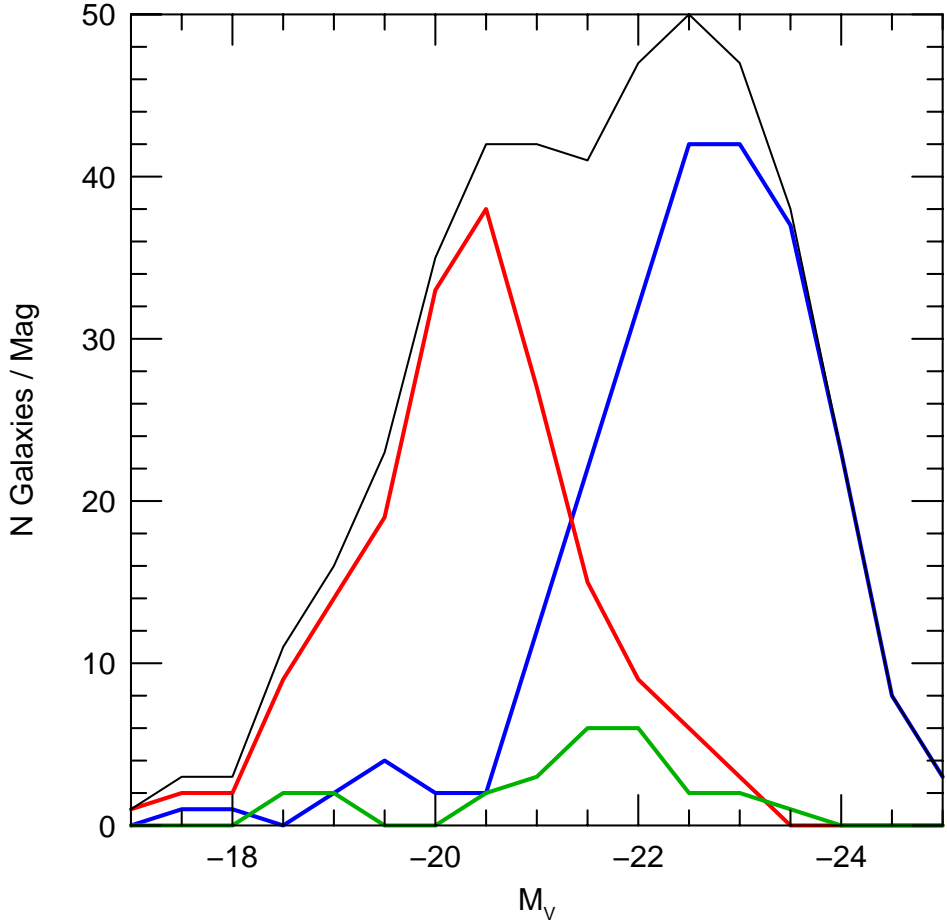}
\caption{Histograms are plotted showing the frequency of core
galaxies (blue line), power-law galaxies (red line), and
intermediate galaxies (green line) as a function of luminosity
in our sample of 219 galaxies.
The thin line shows the total distribution of galaxies with luminosity.
Note that the low-luminosity end of the power-law galaxy
distribution and the high-luminosity end of the core galaxy
distribution are likely to reflect biases with luminosity
in the construction of the sample, but there
are no obvious biases that affect the luminosity range $-22.5\leq M_V\leq-20.5$
over which the frequency of power-law galaxies decreases as that of
the core galaxies increases.}
\label{fig:lum_hist}
\end{figure}

The composition of the sample is well suited to the task of exploring
bimodality, even though it is a composite of several {\it HST} samples.
We are essentially using most of the nearby early-type galaxy observations
that could contribute to such a test.  The overall luminosity distribution
of the sample is shown in Figure \ref{fig:lum_hist}.  The sample is richest
at just the luminosities required to explore the power-law/core transition,
and falls off smoothly at both the bright and faint ends.  The bright end
of the sample is enhanced by the inclusion of BCGs, but as discussed below,
we can test subsamples that exclude the BCGs directly, or all bright
galaxies at large distances.  We can also examine bimodality in a luminosity
slice at the power-law/core transition luminosity, to counter
concerns about the completeness of the sample at fainter and brighter
luminosities.  A separate concern might be the potential effects of different
physical resolution limits within the sample, given that the galaxies
are observed at a variety of distances, particularly if power-law galaxies
are classified as such due to insufficient resolution. \citet{f97} showed
in their sample, however, that upper limits on the break radii for
power-law galaxies fall well below the mean $r_b-L$ relationship at
a given luminosity.  This remains true in
the present sample \citep{l07} for all but the faintest power-law
galaxies.  Power-law galaxies are not simply
poorly resolved core galaxies.

\begin{figure}
\plotone{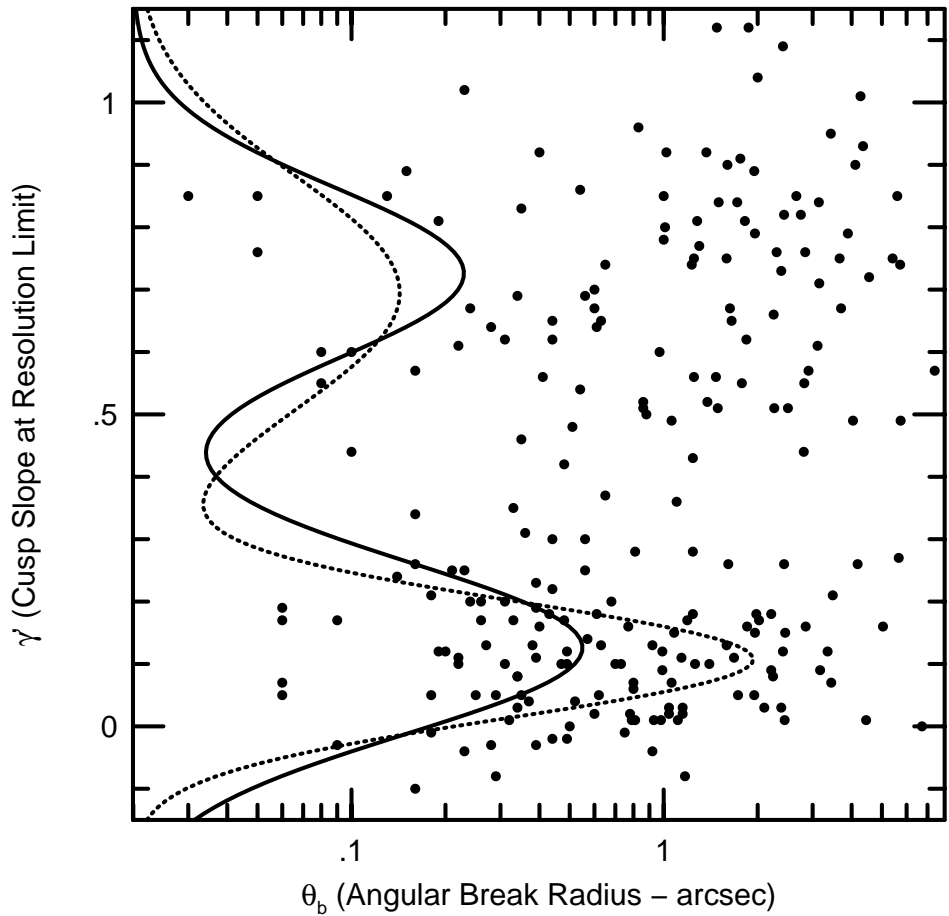}
\caption{The limiting logarithmic slope
$\gamma'$ of the surface photometry profiles of
the sample is plotted as a function of the angular size
of the break radius.  The solid line shows the distribution of
$\gamma'$ modeled as the sum of two gaussians with equal dispersions.
The dashed line shows the distribution when the two gaussians
are allowed to have differing dispersions.  The normalization is
arbitrary, but is the same for both distributions.}
\label{fig:gammap_all}
\end{figure}

\begin{figure}
\plotone{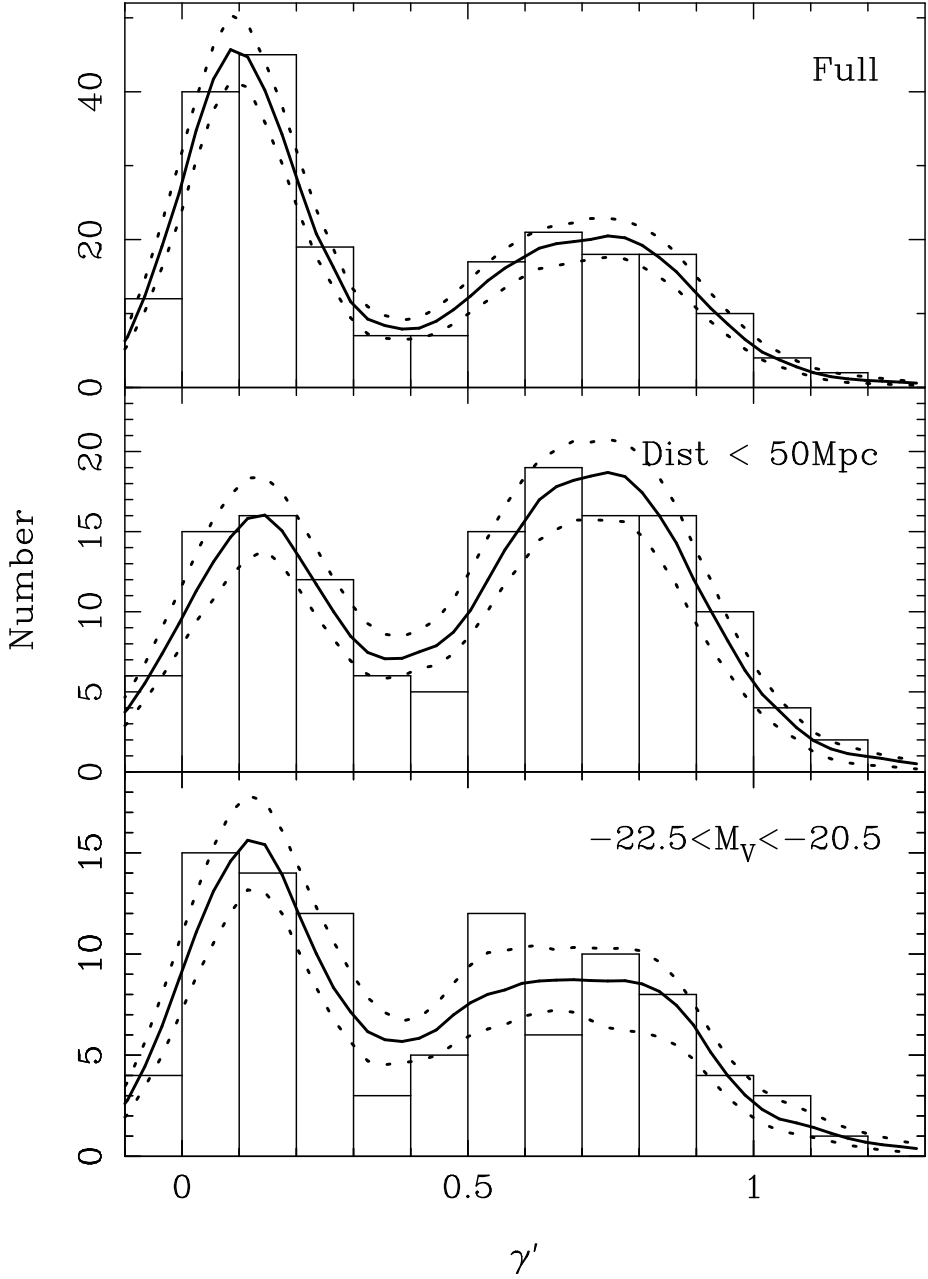}
\caption{Histograms smoothed with an adaptive kernel are shown for three
representations of the sample distribution of $\gamma'.$  The top panel shows
the histogram derived from the entire sample, the second panel just shows
galaxies within 50 Mpc, and the bottom panel just has galaxies with
$-22.5\leq M_V\leq-20.5.$  The dashed lines show the $1\sigma$ confidence
intervals.  The bimodality of all three histograms is highly significant.}
\label{fig:gam_hist}
\end{figure}

Table \ref{tab:cen} lists adopted $\gamma'$ values for the galaxies.
Figure \ref{fig:gammap_all} shows the
new distribution of $\gamma'$ as a function of $r_b$ in angular units.
The overall impression is that the distribution of $\gamma'$ remains
strongly bimodal, even though several galaxies now clearly fall within
$0.3<\gamma'<0.5.$  The bimodality of the complete sample is
strongly evident in the histogram of $\gamma'$ presented in Figure
\ref{fig:gam_hist}.

Rather than just relying on a visual assessment of bimodality, however,
we can test for the statistical likelihood of bimodality using the
KMM test described by \citet{ash}.  The KMM test compares the
likelihood that the frequency distribution of a single parameter
is drawn from the sum of two distinct gaussians versus a single
gaussian distribution.  As \citet{ash} discuss, the most statistically
rigorous test requires that the two gaussians in the bimodal composite
distribution have the same dispersion, but the KMM algorithm
can relax this assumption; we test the $\gamma'$ distribution
both ways.  Table \ref{tab:bimodal} presents the gaussian parameters
describing the distribution of $\gamma'$ shown in Figure \ref{fig:gammap_all};
the total distributions for equal and unequal dispersions are also plotted
in the figure.  The KMM test confirms in both cases that the likelihood
that $\gamma'$ was drawn from a single gaussian rather than a bimodal
distribution is $<10^{-3}.$  In both cases the two gaussians
correspond closely to the core and power-law definitions, with the ``valleys''
in the distributions falling in the region containing the intermediate
galaxies.  The ``power-law gaussian'' is 95\%\ complete for $\gamma'>0.50,$
while the ``core gaussian'' is 95\%\ complete for $\gamma'<0.36.$

A different test for bimodality is provided by
the DIP Test \citep{hnh}, which compares the best-fitting
unimodal distribution function to the empirical distribution function.
The DIP statistic is the maximum difference between these two
distribution functions, which is a measure of the multimodality of the
sample. Monte Carlo simulations drawn from a unimodal distribution
determine the probability of obtaining a measured dip statistic.
The DIP Test allows for greater generality in the possible forms of
the distribution function, and thus is a more conservative statistic
for excluding unimodality.
The value of the DIP statistic for the full sample is 0.047, which
excludes the unimodal hypothesis at 99.9\% significance.

\begin{figure}
\plotone{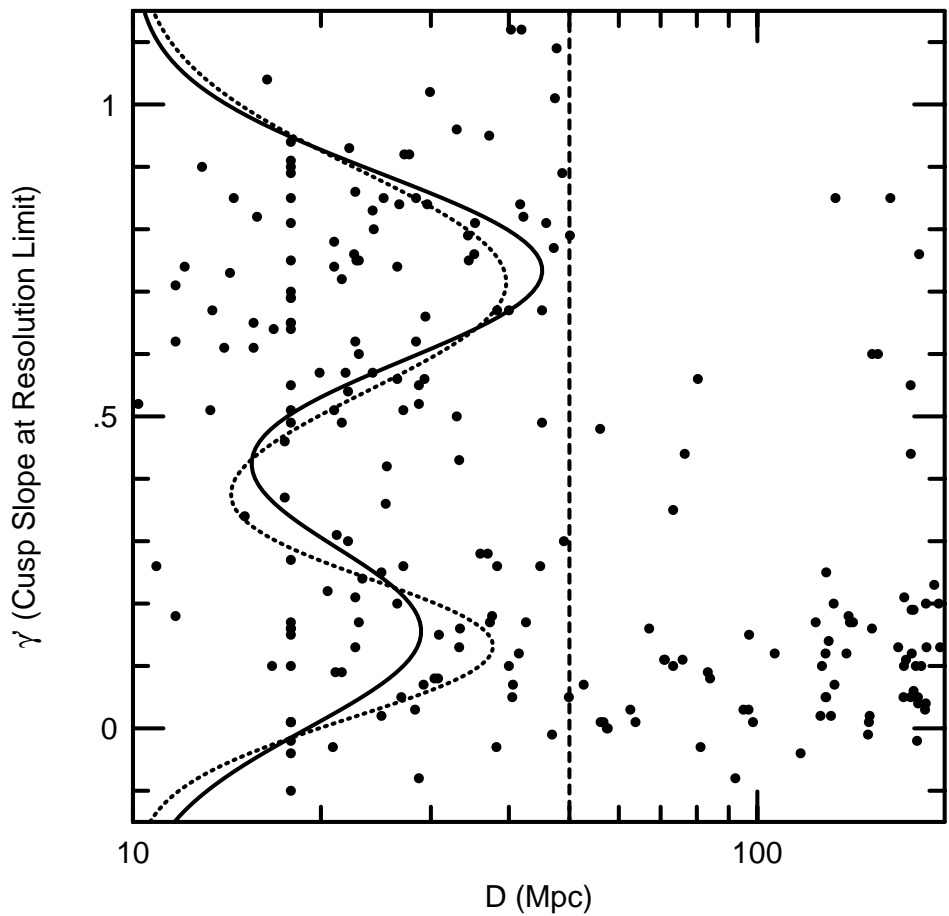}
\caption{The limiting logarithmic slope
$\gamma'$ of the surface photometry profiles of
the sample is plotted as a function of distance.
The solid line shows the distribution of
$\gamma'$ for just those galaxies with $D\leq 50$ Mpc (vertical line)
modeled as the sum of two gaussians with equal dispersions.
The dashed line shows the distribution when the two gaussians
are allowed to have differing dispersions.  The normalization is
arbitrary but is the same for both distributions.}
\label{fig:gammap_d}
\end{figure}

\begin{figure}
\plotone{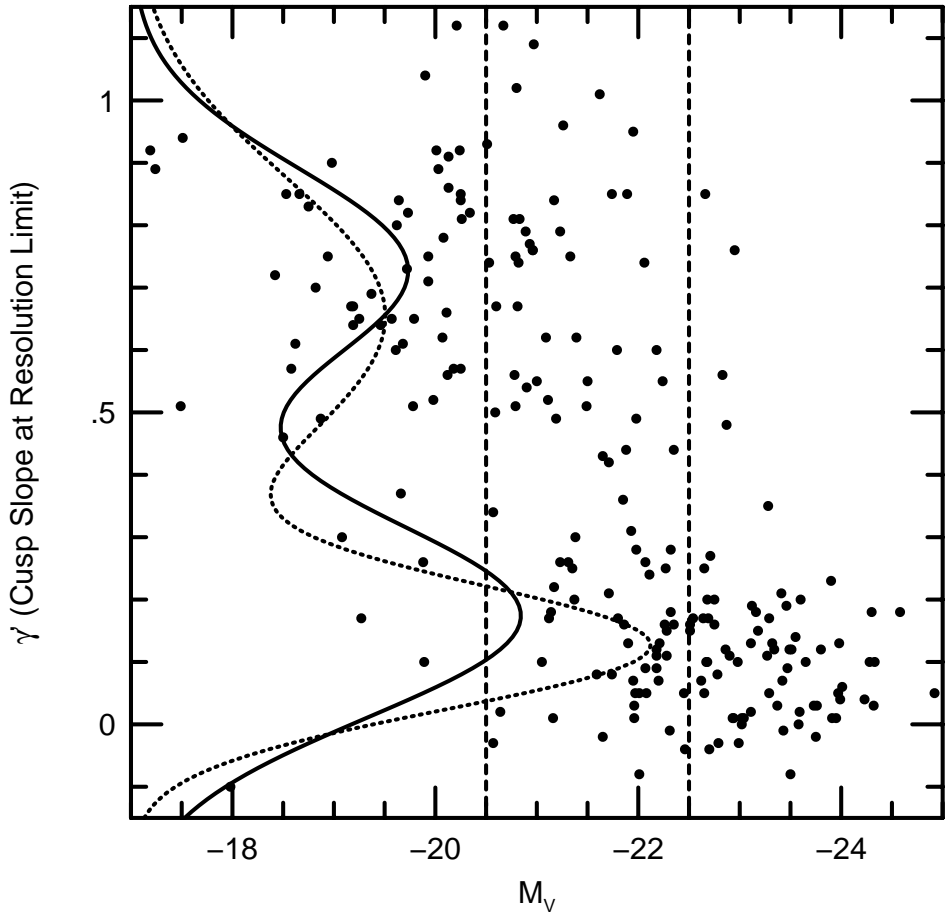}
\caption{The limiting logarithmic-slope
$\gamma'$ of the surface photometry profiles of
the sample is plotted as a function of galaxy luminosity.
The solid line shows the distribution of
$\gamma'$ for just those galaxies with $-20.5\geq M_V>-22.5$ (double
vertical lines)
modeled as the sum of two gaussians with equal dispersions.
The dashed line shows the distribution when the two gaussians
are allowed to have differing dispersions.  The normalization is
arbitrary but is the same for both distributions.}
\label{fig:gammap_mag}
\end{figure}

One caveat is that
the overall distribution of $\gamma'$ is biased
due to the inclusion of the \citet{laine} BCG sample, which greatly
enhances the population of luminous core galaxies.
Figures \ref{fig:gammap_d} and \ref{fig:gammap_mag} plot $\gamma'$ as a
function of distance and luminosity, respectively, allowing some
potential biases in the complete sample to be explored.
Figure \ref{fig:gammap_d}, for example, shows a strong transition
in the population of structural types at $D\approx50$ Mpc.
Since all but two BGC have $D>50$ Mpc, we can easily exclude them
from the sample by considering only those galaxies at $D\leq50$ Mpc;
this cut also excludes several other luminous core galaxies.
The smoothed histogram of the distance-cut subsample shown in
Figure \ref{fig:gam_hist} is clearly bimodal, which is is verified by the
KMM test; again the likelihood that the population was drawn
from a single gaussian is $<10^{-3}.$  The valleys between the
two modal peaks for both the equal and unequal dispersion cases
are not quite as deep (the distributions are plotted in Figure
\ref{fig:gammap_d}), but the location of the two modal peaks for galaxies
with $D\leq50$ Mpc are essentially the same as those for the entire sample.
The value of the DIP statistic for this subsample is 0.050, which
excludes the unimodal hypothesis at 98.2\% significance.

The transition between power-law and core galaxies that
occurs with increasing luminosity is evident in Figure \ref{fig:gammap_mag};
we show the histograms of the structural types with luminosity
in Figure \ref{fig:lum_hist}. 
While the sample has poor completeness at low luminosities,
and the high luminosity population is enhanced by inclusion of the BCG sample,
there are no obvious biases in the middle part of the distribution.
The figure shows that
core and power-law galaxies coexist in roughly equal numbers
in the interval $-22.5< M_V\leq-20.5,$ with the occurrence of power-law
galaxies dropping off rapidly at higher luminosities, and few
core galaxies existing below the low end of this interval.
A complete transition in form occurs over a very small luminosity interval.

The distribution of $\gamma'$ remains bimodal within the luminosity interval
of the power-law/core transition, as is evident in Figure \ref{fig:gam_hist}.
Even though a majority of the intermediate galaxies fall within this
luminosity range, they are still relatively rare.
One cannot describe the population of galaxies
as being drawn from a simple
linear relationship between luminosity and $\gamma'.$
Once again the KMM test rejects the single gaussian at the
$<10^{-3}$ level for representing the distribution of $\gamma'$ for
just the galaxies with $-22.5< M_V\leq-20.5.$  Again, the modal peaks
shown in Table \ref{tab:bimodal} and Figure \ref{fig:gammap_mag}
closely correspond to the core and power-law definitions.
The value of the DIP statistic for this luminosity-cut subsample is 0.053,
which excludes the unimodal hypothesis at 95.5\% significance.

\subsection{Intermediate Galaxies\label{sec:inter}}

The intermediate classification
appears to be highly dependent on resolution, as \citet{rest} noted. 
Indeed, we consider all the galaxies so classified by
Rest et al.\ to be core galaxies, given the
finer resolution limits adopted here for their WFPC2 profiles.
At the same time, we consider NGC 2902, 4482, 5796, and 5898,
which were classified as power-laws by Rest et al., to
be intermediate galaxies.
Likewise, we find the intermediate classification in the
\cite{rav} sample to vary with choice of resolution limit.
In a sense, the tests conducted above might
lead one to relabel the intermediate galaxies
with $0.3<\gamma'<0.5$ as ``ambiguous galaxies.''
The valley in a bimodal distribution is a region where
outliers from each mode coexist and classification of any given
galaxy cannot be done with any degree of reliability.
It is likely that with better resolution some of the present
intermediate galaxies would be classified as core galaxies.
NGC 3585 and 4709, for example, have well-defined breaks, but have $\gamma'$
of 0.31 and 0.32, respectively, falling only just above the core boundary;
they would probably fall below it with better resolution.

At the same time, we are left with the real physical question of
why intermediate galaxies are rare, and the possibility that in
at least some cases their central structure may offer a unique
window on the problem of core formation.
As Rest et al. noted, many intermediates have small $\alpha,$
and do not show as strong breaks as the core galaxies, even
accounting for the higher $\gamma$ values.
In the present sample we have difficulty identifying
any galaxy with $\gamma'\approx0.4$ that shows a sharp break
and well-defined inner cusp --- the best candidates for this
appear to be NGC 2902, 5796, and 5898.
To explore the utility of the intermediate classification
we choose to preserve the
13 intermediate galaxies in the sample as a separate class.
This appears to be sensible based on
the central parameter relationships
presented in \citet{l07}; the intermediate galaxies do not appear
to fall on the relationships defined by core galaxies.

\section{A Non-Parametric Demonstration of Bimodality in Space Density}

Parametric models provide a means to represent data in a
simple way that allows one to determine underlying physical
relationships. However, care has be taken so that unknown biases in
the parameterization do not affect the overall results. This concern
is especially important when studying the luminosity density.
Since it is calculated by deprojecting
the surface brightness profile, any
biases in the brightness profile model will be amplified.
Luminosity density is more fundamental than surface brightness, thus
it is important to verify that the bimodality of cusp slopes in
projection truly reflects the intrinsic distribution of space-density
slopes.  We therefore bolster the result in the previous section
with a non-parametric approach.

The uncertainties for non-parametric models will generally be larger than those
resulting from parameter fits, and should provide a more realistic
estimate. The comparison between the two approaches was done in
\citet{g96}, where we found that the Nuker law provides an accurate
estimate of the surface brightness profile in the central regions of
early-type galaxies. While detailed analysis of the luminosity
density, including dynamical studies, should incorporate a
non-parametric approach, use of the Nuker law facilitates comparison
with theoretical models and parameter correlation studies. Therefore,
both are appropriate.

We follow the same procedures as in \citet{g96} to estimate the
surface brightness and luminosity density profiles.  The technique
represents the measured brightness profiles with smoothing splines,
and then inverts the splines into density space.  Monte Carlo
simulations are used to establish the errors in the final density
profiles.  As the present analysis is identical to that of \cite{g96},
we refer the reader to that paper for a complete description.

\begin{figure}
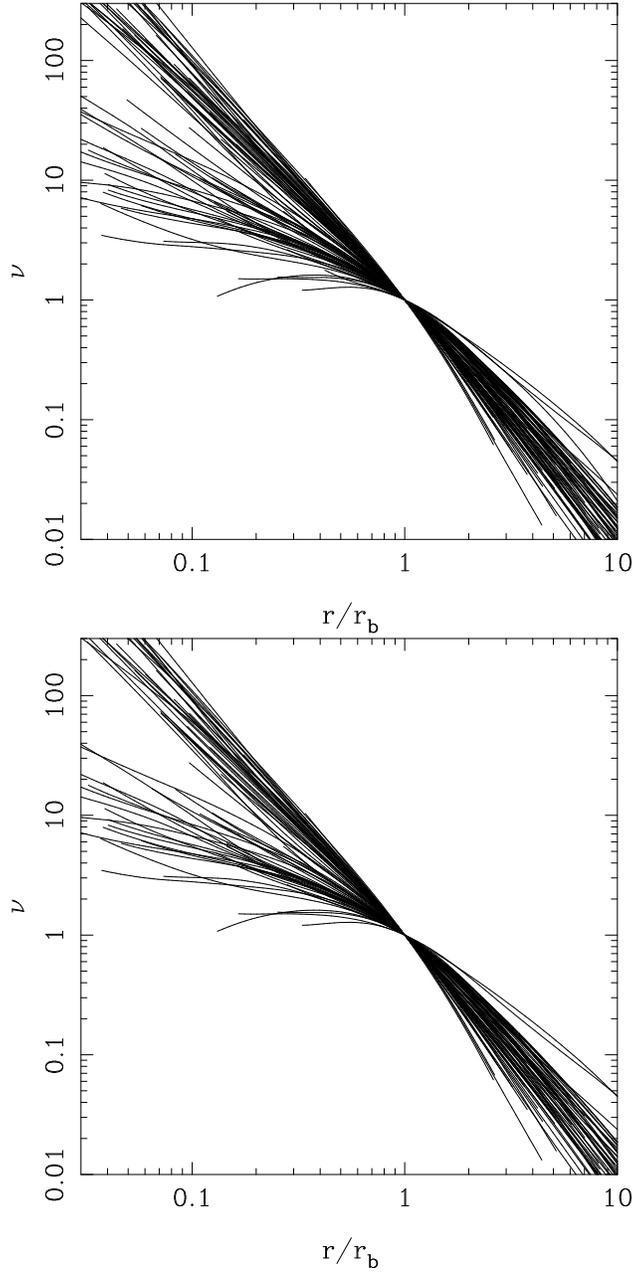

\centering
\includegraphics[scale=0.5]{f06a.ps}
\includegraphics[scale=0.5]{f06b.ps}
\caption{Luminosity density profiles are shown for the subset of galaxies for
which WPFC2 brightness profiles are available (as opposed to galaxies for
which we have only the Nuker-law parameters).
The \citet{laine} BCG sample is also excluded.
The radial scale is normalized by Nuker law break-radius, but the real
measured brightness profiles were used rather than the Nuker-law fits.
The slopes of the profiles for $r<r_b$ are strongly bimodal.  The
lower graph excludes galaxies in \citet{l05} with dust strength greater
than 1.0.}
\label{fig:den_pro}
\end{figure}

Figure \ref{fig:den_pro} shows the luminosity density profiles
normalized in both radius and luminosity density of all galaxies for
which we had the measured isophotal parameters (the \citealt{l95},
\citealt{l05}, and \citealt{laine} samples). The normalization is at
the radius that shows the largest logarithmic second derivative in the surface
brightness profile; this point corresponds to the break radius in a
Nuker law fit, but again it is determined in this context non-parametrically.
This plot is analogous to Figure 3 in \citet{g96};
however, the \citet{g96} sample only contained 42 galaxies,
whereas the present plot is based on 85 galaxies.
The two main points evident from both plots are that nearly all galaxies show a
central cusp in the luminosity density and that galaxies separate into
two types: strong and weak cusp (i.e., power-law and core galaxies).

Again, Figure \ref{fig:den_pro} shows only a subset of the composite sample
constructed in this paper, comprising the profiles of \citet{l95},
\citet{f97}, and \citet{l05} for which we have the directly measured
WFPC1 or WFPC2 surface photometry. We have excluded the BCGs from
\citet{laine}, but their inclusion mainly adds core galaxies.
The sample in Figure \ref{fig:den_pro} contains more galaxies in the transition
region between the core and power-law types than did the
sample in \citet{g96}, but the
statistical difference between the two profile classifications remains
as strong. We have checked whether dust has an effect by excluding
galaxies with central dust strengths greater than 1.0 \citep{l05}. The
resulting composite plot (the lower panel in  Figure \ref{fig:den_pro})
shows even cleaner bimodality in the luminosity density compilation.

\begin{figure}
\centering
\includegraphics[scale=0.7]{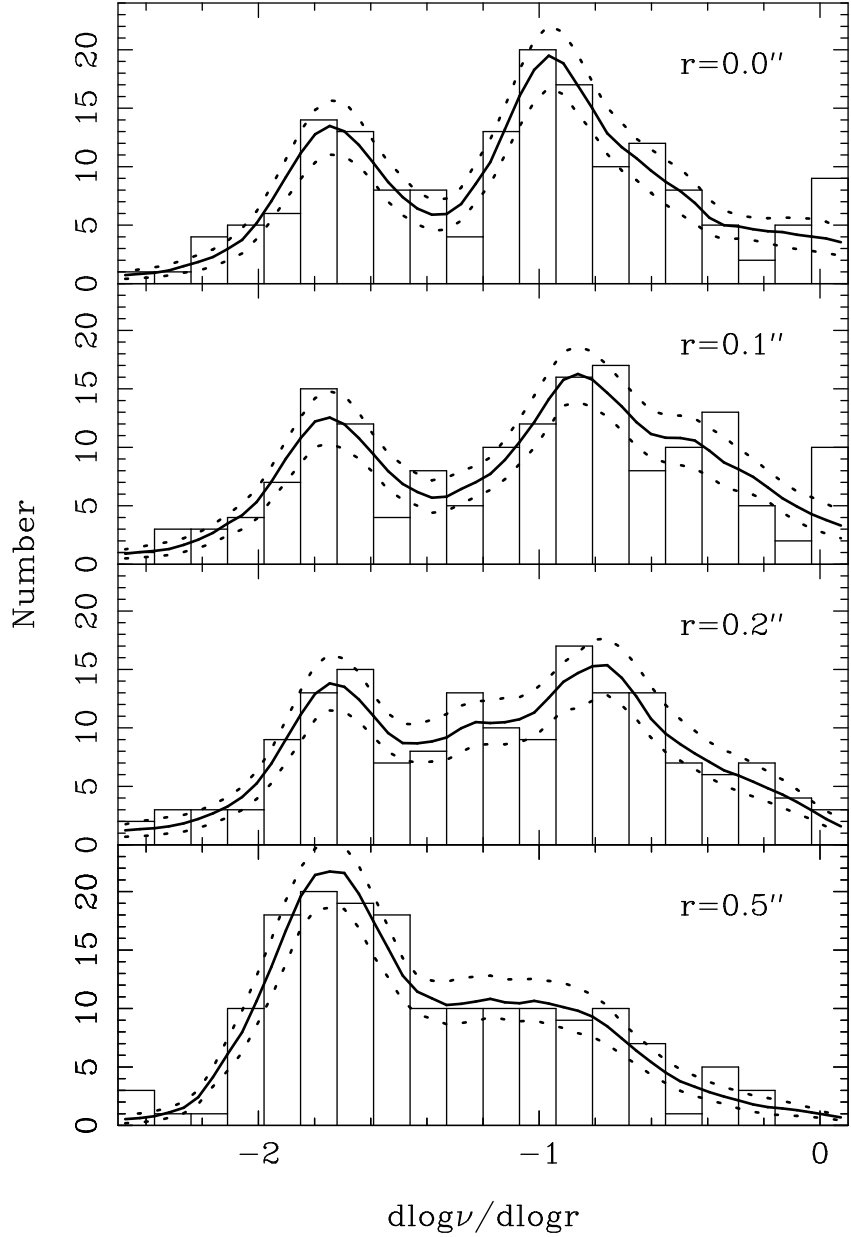}
\caption{Histograms of the limiting slope of the luminosity density profiles
are shown for various limiting radii.  The bimodality of the histograms
with the limiting $r\leq0\asec1$ is highly significant.}
\label{fig:den_hist}
\end{figure}

We have measured the value of the density logarithmic slope at various radii
to determine the significance of the bimodality.
We plot histograms and density estimates of the slopes of luminosity
density measured at various radii in Figure \ref{fig:den_hist}. For
these plots, we now include all 166 galaxies for which we have
measured profiles. Overplotted on the histograms are adaptive kernel
density estimates and their 68\%\ confidence bands \citep{silver}. As
one goes to larger radii to measure the slope, the significance of the
two peaks lessens, and beyond $r=0\asec5$
we find a unimodal distribution of
slopes. There are many statistical tests to determine the significance
of bimodality, and most give very similar results.  Using the DIP test
For the four radial limits presented ($r=0\asec0,\ 0\asec1,\ 0\asec2,$
and $0\asec5$), we find significance values of 0.997, 0.98, 0.80, and 0.40.
The hypothesis that the central slopes are unimodal at the
{\it HST} resolution limit is highly excluded, with significance
very similar to what we find from the Nuker fits using the KMM algorithm.

\section{A Comparison to the ACS Virgo Cluster Survey Profiles}

The analysis in the previous sections conflicts strongly with the
\citet{lf} conclusion that a bimodal cusp distribution is ``untenable.''
\citet{lf} derived surface photometry profiles of 100 early-type
galaxies in the Virgo cluster as part of their ACS Virgo cluster survey.
Their methodology differs from ours in a variety of ways.  They used the
ACS/WFC, observed in the SDSS {\it g} and {\it z} filters,
and measured surface photometry as a function of isophote
geometric mean radius; however, all of these differences are trivial.
A more significant difference is that the \citet{lf} sample extends to much
fainter systems than we have studied, while we have much richer coverage
of bright ($M_V<-19$) systems where the core/power-law transition occurs.
The relationship of the central structure of the faint dwarf elliptical
galaxies studied in the VCS to that of brighter ellipticals is an
important question, but as we discuss later, it is also one that can be
cleanly separated from the physical interpretation of bimodality
that we have advanced in \citet{f97}, namely that at $M_V\approx-21,$
the central structure of elliptical galaxies makes a rapid transition
in form with increasing luminosity,
which is well correlated to other physical properties of the systems.
The VCS analysis also finds that
core galaxies appear above this luminosity,
but their physical picture of less luminous galaxies
is markedly different than ours.

The divergence between us and \citet{lf}
appears to be mainly due to differences in our respective models used
to describe the inner surface brightness profiles of ellipticals.
We fit Nuker laws directly to profiles measured from PSF-deconvolved images.
\citet{lf} use three different forms, but in all cases
are fitting PSF-convolved models to the images as observed.
One model uses the \citet{sersic} form alone,
\begin{equation}
I(r)=I_e\exp\left[-b_n\left(\left({r\over r_e}\right)^{1/n}-1\right)\right],
\end{equation}
while a second includes an additional King-model nucleus added to this form.
For their core galaxies, \citet{lf} use
Core-\Ser models \citep{grcs}, but with $\alpha=\infty$,
which mean the galaxies are actually fitted with two different forms
that intersect at the break radius,
\begin{eqnarray}
I(r)&=&I_b\left(r_b/r\right)^\gamma,\qquad r\leq r_b,\nonumber \\
&=&I_b\exp\left[b_n\left(\left(r_b/r_e\right)^{1/n}
-\left(r/r_e\right)^{1/n}\right)\right],\qquad r>r_b.
\end{eqnarray}
While there are {\it a priori} reasons for why one might prefer one approach
to the other, the only real issue is which approach has better
fidelity to the data.  A direct comparison of the VCS
models and Nuker law fits to deconvolved brightness profiles shows that the
Nuker models are superior representations of the observations.
Further, the differences between these models and our data are always
echoed in the model residuals presented by \citet{lf} for their own data
(except where the VCS model includes a large nuclear component,
which absorbs the residuals). 
We conclude that our different results for the same galaxies are not
due to differences in the basic image data, nor even the use of deconvolution
versus PSF-convolved model fitting (although recognizing
subtle systematic failures of the models is considerably more difficult
in the observed domain), but in the accuracy and appropriateness
of the different profile parameterizations used.

\subsection{A Comparison of VCS and Nuker $\gamma'$ Values for Galaxies
in Common}

The present sample has 27 galaxies in common with the 100 VCS galaxies.
As these are preferentially brighter systems,
the overlap between the two samples in the luminosity range
associated with the transition between core and power-law galaxies is
much more significant; the great majority of the
common systems are brighter than the median VCS galaxy luminosity.
Of the 27 galaxies, 22 were observed by us in the course of our own
{\it HST} observational programs.  The majority of these were imaged
with WFPC2, but a substantial minority were obtained with WFPC1.
As discussed in \cite{l05}, deconvolved profiles obtained with
both cameras agree extremely well for $r>0\asec1;$ we show below
that both cameras also agree as well with deconvolved ACS/WFC
observations.  Differing resolution between the three cameras is not
a factor that contributes to the differences between the present
and VCS results.

Figure \ref{fig:gam_diff} plots $\gamma'$ values from the present
sample against those of the VCS sample for the 27 galaxies in common.
The agreement between the samples is poor, but there are clear
trends in the pattern of the differences.
The VCS $\gamma'$ values for the core galaxies are always
systematically higher than our values, while the VCS $\gamma'$
values for the power-law galaxies are uniformly lower.  The
dynamic range of the VCS $\gamma'$ values is thus overall smaller than ours,
and many of the VCS $\gamma'$ values place the galaxies in
the intermediate zone between cores and power-laws.
In a number of cases, often where \citet{lf} claimed the existence
of large nuclei, the agreement between the two samples is especially poor,
with the VCS $\gamma'$ values falling in the range of core galaxies
for systems that we had classified as power-laws.

\subsection{A Comparison of Deconvolved ACS, WFPC1, and WFPC2 Profiles}

The ACS/WFC imagery used by the
VCS study and the WFPC2 and WFPC1 imagery that dominate the present
sample provide essentially equivalent information.
A comparison of WFPC2/PC1 and ACS/WFC PSFs shows that WFPC2 actually provides
significantly better angular resolution (Figure \ref{fig:psf}).
The FWHM of the WFPC2/PC1 F555W is $0\asec061,$ while that of the
ACS/WFC F475W PSF is $0\asec092,$ or 50\%\ larger.
The use of {\it Drizzle} \citep{driz} to rectify the ACS images slightly
degrades its resolution further.
The present sample has 49 galaxies at Virgo distance or closer,
and a substantial number that are no more than 50\%\ more distant;
both the present and VCS samples are probing the same physical scales
in the galaxies.

Regardless of the resolution difference between WFPC2 and ACS,
PSF-deconvolution provides a model-independent approach to correct
{\it HST} imagery for the ``wings'' of the PSF outside the {\it HST}
diffraction limit and to isolate the profile analysis from the residual blurring
interior to the diffraction limit that remains even after deconvolution.
The deconvolution tests presented in \citet{l92b}, \citet{l95}, \citet{l98},
\citet{rest}, and \citet{l05} have established that this methodology
works well for {\it HST} studies of brightness profiles; they also
yield well-understood resolution limits.
The ACS images can be deconvolved as well to correct for differences
between the ACS and WFPC2 PSFs outside the resolution limit.
We demonstrate this with direct comparisons of deconvolved ACS and WFPC2
or WFPC1 profiles for several galaxies in common.

Figure \ref{fig:acsdecon} compares profiles measured from PSF-deconvolved
ACS F475W ``drizzled" images to those obtained from deconvolved
WFPC1 F555W images \citep{l95} for eight galaxies in common,
and from deconvolved WFPC2 F555W images \citep{l05} for two galaxies in common.
We present more extensive comparisons between ACS and WFPC1 than between
ACS and WFPC2 because this is the more challenging test;
the WFPC1 imagery was blurred by a strongly aberrated PSF.  Further,
most of the power-law galaxies in common with the VCS sample were observed
with WFPC1 rather than WFPC2; the steep central cusps in power-laws
present a greater challenge for deconvolution than do core galaxies.

The agreement between all three cameras in all cases is excellent, with
differences between the profiles for $r>0\asec1$ limited to only a few percent.
This is consistent with the expectations from the simulations and
tests quoted above, which show that WFPC1 deconvolutions are largely
accurate into $r\approx0\asec1$ (with the caveat noted in \citealt{l05} that
WFPC1 deconvolutions may be less accurate at this radius if a strong
nuclear point source is present).
ACS/WFC has superior resolution for $r<0\asec1,$ as evidenced by
the systematically higher central surface brightnesses recovered interior
to this radius.  Since Figure \ref{fig:acsdecon} shows nearly all
of the 10 WFPC1 overlap galaxies\footnote{We exclude NGC 4434,
for which the archived drizzled-image has
a strong artifact in the galaxy center, and NGC 4486, which is a core galaxy,
and for which \citet{l92a} removed the nuclear non-thermal source
from the WFPC1 images.},
it demonstrates directly that the use of WFPC1 profiles in the present
sample is not a factor in the strongly different $\gamma'$ values derived from
the Nuker and VCS models.

The two galaxies selected for the WFPC2/ACS comparisons are the most
centrally compact galaxies observed with WFPC2 in the overlap sample,
and thus are the most sensitive to differences in resolution.
The WFPC2 profiles show superior resolution to
ACS/WFC for $r<0\asec1,$ which is expected, given the sharper WFPC2 PSF.
Again, however, none of the differences between the VCS results and ours depends
on the exquisite treatment of details near the {\it HST} diffraction limit,
as we will show in the next section.

\subsection{Comparison of Nuker Laws and VCS Profile Models
for Galaxies in Common}

\begin{figure}
\plotone{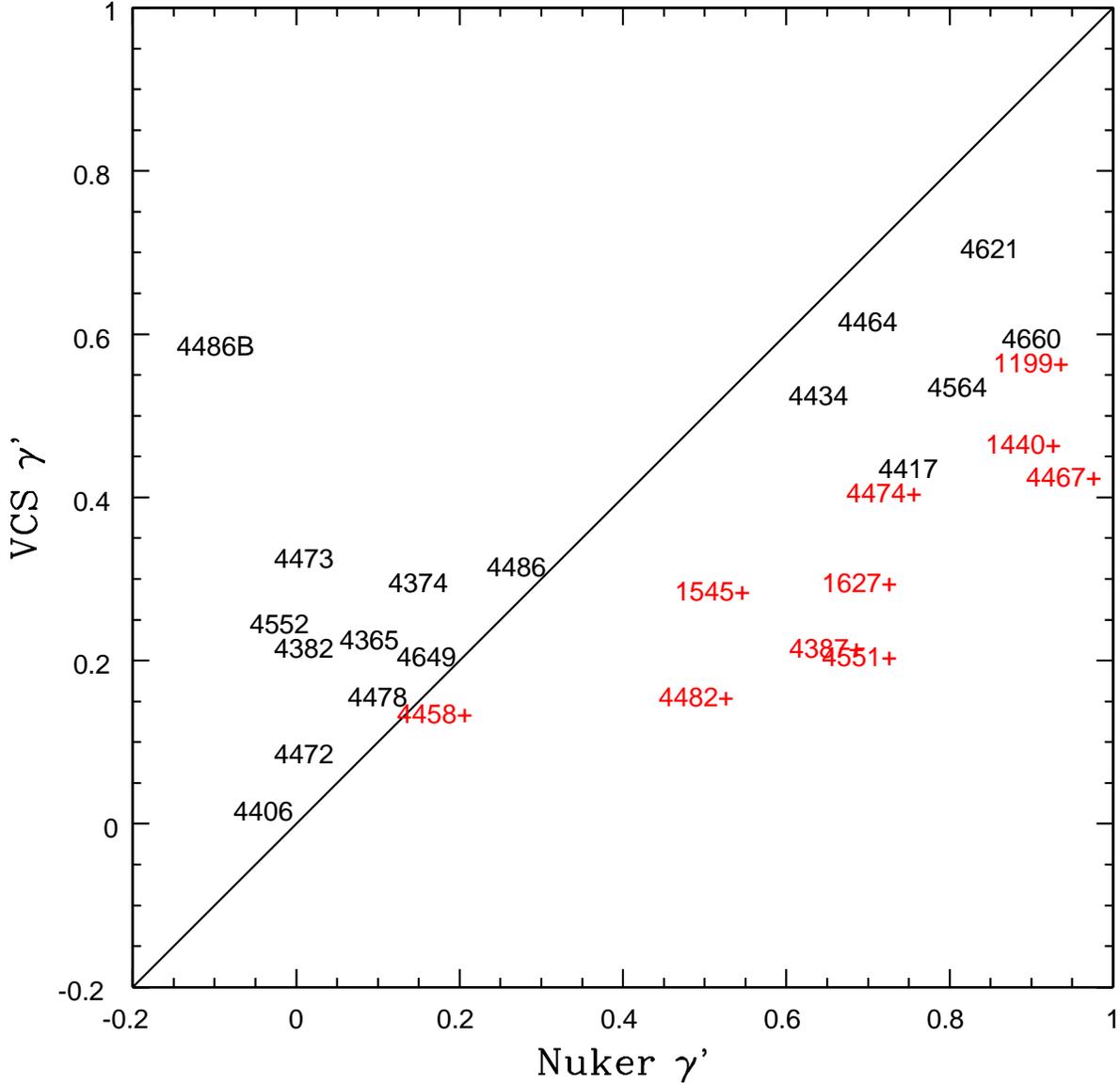}
\caption{The limiting logarithmic slope $\gamma'$ of the galaxy
brightness profiles as the {\it HST} resolution limit is approached
is plotted for 27 galaxies in common between the present and
\citet{lf} VCS samples.  The horizontal axis gives $\gamma'$ estimated
from Nuker law fits.  The limiting resolution is typically between
$0\asec04$ and $0\asec1,$ depending on the source material.  The
vertical axis gives $\gamma'$ estimated
at $r=0\asec1$ from the VCS \Ser or Core-\Ser models.  Galaxies
plotted in red and marked with ``$+$'' are systems in which \citet{lf}
fitted a \Ser model plus a central nuclear component.}
\label{fig:gam_diff}
\end{figure}

\begin{figure}
\plotone{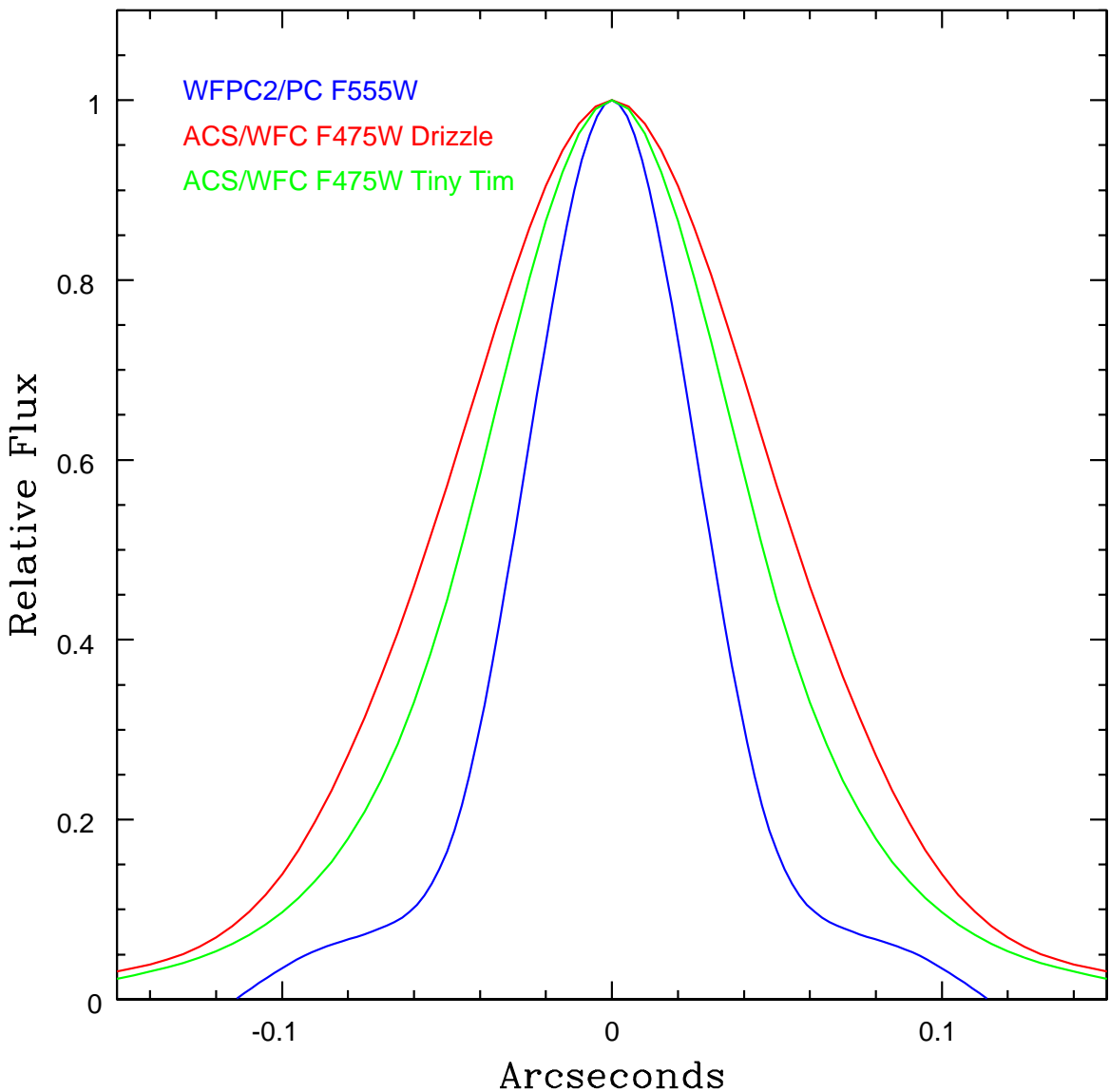}
\caption{The widths of the WFPC2/PC F555W (blue) and ACS/WFC F475W PSFs
(green and red) are compared.
The highest resolution material in the present sample uses the former
filter/instrument combination, while the \citet{lf} VCS program uses the
latter.  The two traces for the ACS PSF are from the {\it Tiny Tim}
calculated PSF (green), and from a star in a ``drizzled'' image (red),
which slightly degrades the native PSF.  The WFPC2 PSF is significantly
sharper.}
\label{fig:psf}
\end{figure}

\begin{figure}
\plotone{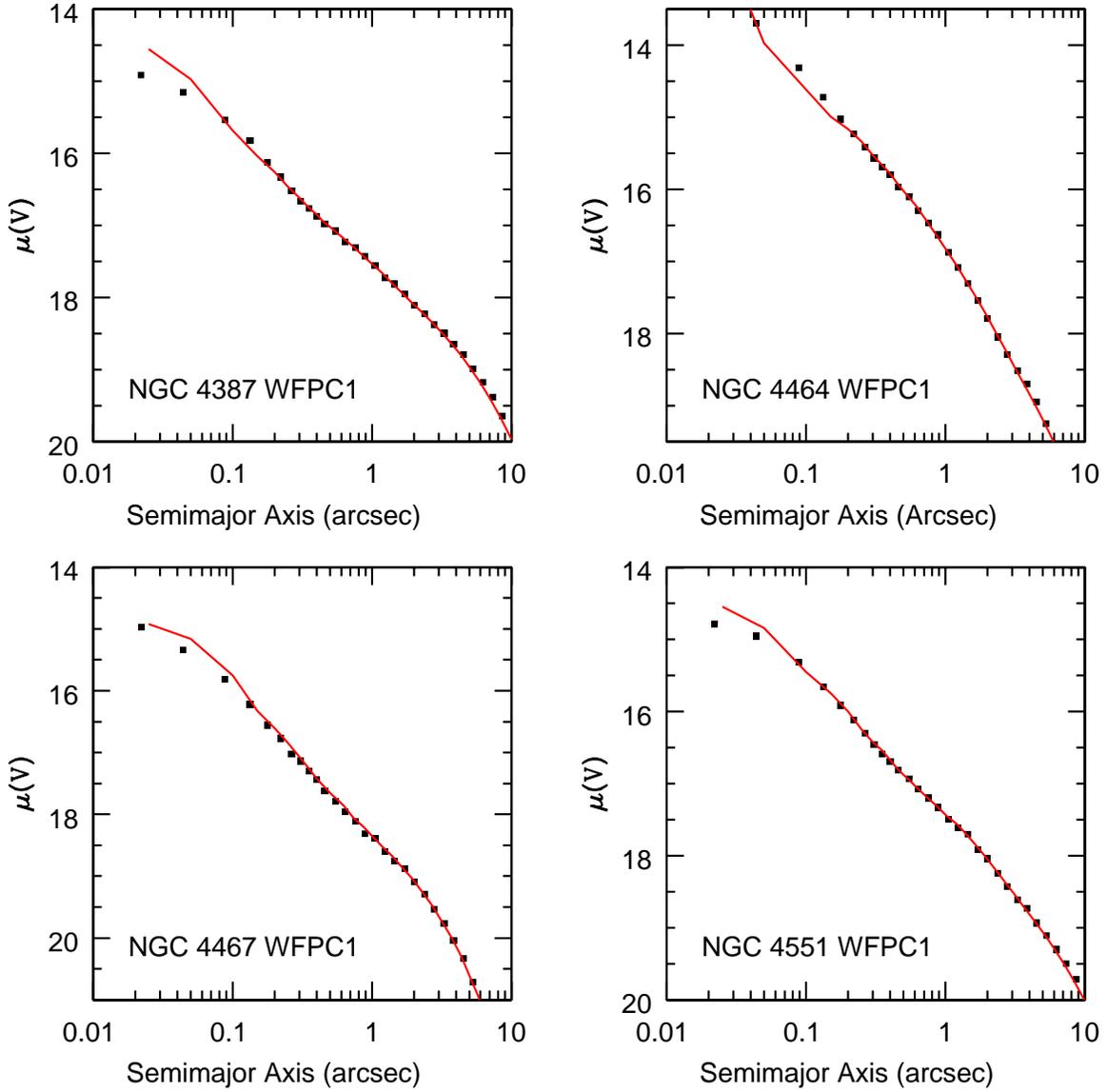}
\caption{Deconvolved brightness profiles obtained with ACS/WFC
are compared to WFPC1 or WFPC2 profiles for 10 galaxies
in common the to present and \citet{lf} VCS sample.  Profiles measured
from the deconvolved and drizzled ACS/WFC F475W images are shown in red
as compared to deconvolved F555W photometry from WFPC1 \citep{l95} or WFPC2
\citep{l05}.  The first eight galaxies were observed with WFPC1.  NCG 4478 and
4486B were observed with WFPC2 (the NGC 4478 images were dithered to
create an image with sub-sampled pixels). Deconvolved profiles
from ACS, WFPC2, and WPFC1 images all agree well into $r\approx0\asec1;$
interior to this radius the comparisons show that ACS/WFC has superior
resolution to WFPC1, but that WFPC2 is superior to ACS.}
\label{fig:acsdecon} 
\end{figure}

The VCS and present $\gamma'$ values both come from models of the
galaxy intrinsic light distributions.  Understanding the differences
between the $\gamma'$ values shown in Figure \ref{fig:gam_diff} requires
comparing the VCS and Nuker intrinsic models to the deconvolved surface
brightness profiles, which are themselves a non-parametric representation
of the intrinsic light distributions of the galaxies.
The concordance between ACS and WFPC2 or WFPC1 deconvolved profiles
further means
that we can directly compare the VCS intrinsic profile models (the
adopted representation of the profiles {\it prior} to PSF convolution)
to the deconvolved WFPC2 or WFPC1 profiles, as well as comparing the
Nuker models to deconvolved ACS/WFC profiles.
In practice, we can select the camera that provides the highest spatial
resolution for a given galaxy, thus comparing both the Nuker and VCS models
to the best deconvolved profile of the galaxy.  In general, this means using
WFPC2 profiles as the first choice, with the ACS as the second choice
(WFPC1 profiles will still be used for the two exceptions noted above).

The disagreement of $\gamma'$ values shown in Figure \ref{fig:gam_diff}
can be traced to differences between the VCS and Nuker profile models.
We show this on a galaxy
by galaxy basis in Figures \ref{fig:acscomp1} to \ref{fig:acscomp3}
by comparing the
published {\it intrinsic} VCS {\it g}-band models and
Nuker models to WFPC2, ACS, and WFPC1 deconvolved isophotal brightnesses.
The measured brightness values are presented twice.  The lower trace
has the isophotal scale converted to mean radii to accommodate the
VCS models, which are normalized to the surface brightness measures
at $r>2''$ to account for the color offset between the WFPC1/2 F555W
and ACS F475W filters.  The upper traces are the same data
offset by 0.5 mag and fitted with the Nuker models using our preferred
semi-major axis radial scale (these plots were shown earlier in
either \citealt{byun} or \citealt{l05}).  Note that the Nuker laws
often are not fitted all the way into $r=0.$ The limits shown
are those we have adopted to avoid central nuclei, and so on, and
when larger than the general resolution limits presented in
$\S\ref{sec:samp}$ are always the scale
at which we measured $\gamma'.$  Because
we are working in the deconvolved domain, we can exclude the points
interior to some radial limit from the fit.  In contrast, models
fitted to PSF-convolved images
must always account for the flux at $r=0,$ given its contribution
to larger radii in the observed domain.  We thus show the VCS models
into the smallest radii for which isophotal parameters have been measured.

We do not present any quantitative measure of how well the Nuker
versus VCS models fit the data --- simple examination of Figures
\ref{fig:acscomp1} to \ref{fig:acscomp3} shows that the Nuker fits
provide a better representation
of the data in all 22 galaxies.  The differences between
the Nuker and VCS models can be better understood
by grouping the galaxies into three sets:
galaxies classified having core type profiles in this paper, galaxies
classified as power-law profiles in this paper for which \citet{lf}
did not include a nuclear component, and power-laws
for which they did include a nucleus.
 
\subsubsection{VCS Models of Core Galaxies}

\citet{lf} described the profiles of
the most luminous core galaxies (NGC 4365, 4382, 4406, 4472, 4486,
4552, and 4649) with the \citet{grcs}
Core-\Ser model, which features a power-law cusp at small radii
transitioning to a \Ser model at large radii.  As with the Nuker law,
the radial extent or smoothness of the transition is moderated by the
$\alpha$ parameter.  \citet{lf}, however set $\alpha=\infty,$
causing a sharp unphysical slope discontinuity
the break radius, $r_b,$ which sets the radius
of transition between the inner cusp and outer \Ser law.\footnote{While
the $r_b$ parameter in the Core-\Ser law serves an analogous role
to the Nuker break radius, $r_b,$ they are not mathematically equivalent
and cannot be compared directly.  The Nuker $r_b$ occurs at the point
of maximum profile curvature (in log-log coordinates),
while for the Core-\Ser form this point is an exceedingly complex
expression involving all parameters except the photometric scale.}
Setting $\alpha=\infty$ also forces a cusp of constant
logarithmic slope to be fitted for all $r<r_b.$

As is clear in Figure \ref{fig:acscomp1}, the inner
profiles of the core galaxies curve downward as they make a smooth
transition from the shallow central cusps to the steeper envelopes.
This transition is well described by the Nuker law; indeed
\citet{l95} emphasized the need for allowing $\alpha$ to be a free parameter.
Force-fitting a cusp with constant logarithmic slope for
$r<r_b$ yields a $\gamma'$
value that represents the average slope interior to $r_b.$  As the
slope typically continues to decrease as $r\rightarrow0,$ the
true $\gamma'$ at the resolution limit (or terminus of the Nuker-law
fit) will be systematically smaller, explaining the systematic
effect seen in Figure \ref{fig:gam_diff}.

The departures of the VCS models from the WFPC2 data appear to be
completely consistent their residuals from the original ACS/WFC data as well.
The patterns produced by assuming the inner cusp to have constant slope are
readily evident in the residual traces presented in Figure 103 of \citet{lf}.
Because that figure only presents residuals in the observed
domain, the systematic trends are more subtle than those evident in Figure
\ref{fig:acscomp1}; however, one always sees a sharp bump at $r=r_b,$
with a nonzero residual slope at smaller radii.

In the cases of NGC 4473, 4478, and 4486B,
\cite{lf} fitted \Ser models to galaxies that we classified as
core systems.  NGC 4473 is well described by a Nuker law
with a small $\alpha$ parameter.  The \Ser model in contrast clearly
over estimates the central $\gamma';$ again the pattern of residuals that we
see here is obvious in VCS Figure 103 as well.
This is also true for NGC 4486B.
The extremely large residuals from the \Ser fit for NGC 4478
seen in Figure \ref{fig:acscomp1} are again echoed in VCS Figure 103.
While we concluded that NGC 4478 has a small nucleus
\citep{l05}, no such component was included in the VCS model fits.
Lastly, one low-luminosity core galaxy, NGC 4458, was fitted with
a highly luminous and spatially extended nucleus by \citet{lf};
it will be discussed in the context of other galaxies with large
assumed VCS nuclei below.

We conclude that none of the present core galaxies observed in the VCS
are well matched by the VCS models as the resolution limit is approached.
The VCS choice of a cusp with a constant inner slope or fitting a 
\Ser model to galaxies with small cores always results in higher $\gamma'$
values than result from the Nuker parameters. 

\subsubsection{VCS \Ser Models of Power-law Galaxies}

NGC 4434, 4464, 4621, and 4660 are galaxies that we have classified as
power-laws that \citet{lf} fitted with pure \Ser laws.
The VCS models for all four galaxies fall below the data
as $r\rightarrow0$ (Figure \ref{fig:acscomp2}).
Again, this behavior is echoed in VCS Figure 103, which shows that
the \Ser fits fall below the {\it observed domain} ACS VCS data
for $r<0\asec3$ in NGC 4464, 4621,
and 4660, and $r<0\asec15$ in NGC 4434.
\citet{lf} note that the ACS F475W
images of both NGC 4621 and 4660 are actually centrally saturated.
This is problematic as information about the central structure is lost
once saturation occurs, but it still must be accounted
for in the profile models, given that light from the center is also
conveyed to larger radii by the PSF.
In contrast the Nuker models nicely match all four galaxies,
In the case of NGC 4464 it is especially notable
that the Nuker profile, which was derived from WFPC1 imagery, has better
fidelity to the ACS profile than the VCS model derived directly
from the same data.

The plot of Nuker $\gamma'$ versus VCS $\gamma'$ in Figure \ref{fig:gam_diff}
contains two more power-law galaxies, NGC 4417 and 4564, for which we do not
have our own WFPC2 or WFPC1 profiles, but again the residuals in VCS Figure 103
show that their \Ser models fall well below the ACS data as
$r\rightarrow0.$  This appears to be a generic feature of nearly all the
VCS galaxies modelled as non-nucleated \Ser profiles, looking at the VCS
galaxies for which we do not have our own data.  NGC 4483, 4528, and 4570
are additional examples of where the departure of the VCS \Ser models
below their own data is especially large as the {\it HST} resolution
limit is approached.

In summary, none of the present power-law galaxies are well matched by the
VCS pure \Ser models as the resolution limit is approached.
The \Ser models always fall below the data at the center, resulting
in smaller $\gamma'$ values than are inferred from the Nuker parameters.

\subsubsection{VCS \Ser $+$ Nucleus Models of Power-law Galaxies}

The set of galaxies in Figure \ref{fig:acscomp3}, NGC 4387,
4467, 4551, VCC 1199, 1440, 1545, and 1627, are power-law galaxies
to which \citet{lf} fitted \Ser laws plus luminous extended nuclear components.
Deconvolved ACS images provide all the profiles for this set of comparisons.
\citet{l95} identified nuclei in all these galaxies as well
(with the exception of NGC 4467), albeit ones of markedly lower luminosity
and extent than those presented by \citet{cote}.  It is evident
from examining the profile fits in Figure \ref{fig:acscomp2} that the
Nuker laws can accurately describe the deconvolved
ACS profiles of these galaxies into small radii,
even though all the Nuker models in this subsection
were derived from WFPC1 imagery.

The VCS \Ser $+$ Nucleus Models {\it in toto} are also excellent models of the
data and thus are physically plausible representations, but the total
quality of the fit is not relevant to the accuracy of $\gamma'.$
The VCS \Ser components {\it alone}, from which $\gamma'$ is measured,
always fall below the data well before the resolution limit is reached
(Figure \ref{fig:acscomp3}).  The implied nuclei often greatly dominate
the inner regions where {\it HST} provides unique resolution,
The VCS $\gamma'$ values for such systems are thus inward extrapolations
evaluated at radii that in some cases are over an order
of magnitude smaller than radii at which the \Ser profile actually dominates.
In contrast, the $\gamma'$ values from the Nuker laws always
reflect the end point of the radial interval over which the fit was performed.
We see no reason why an extrapolated $\gamma'$ measurement from a \Ser law
fitted to the envelope should be used in preference to the local $\gamma'$
at the resolution limit provided by the Nuker law.

\citet{cote} point out that definition of what constitutes a nucleus can
depend on what the underlying assumed profile is.
A corollary of this, however, is that nuclei so identified will not be unique.
The prior existence of the Nuker fits, which also assume small nuclei for nearly
all for the galaxies listed above, establishes this.
It is worth noting that \Ser models had not been tested for use to describe
the inner portions of {\it HST} brightness profiles prior to the VCS work
for more than a few galaxies.  The VCS work, however, takes it as an
{\it a priori} assumption rather than as a hypothesis
that the envelopes of galaxies, which is where the
\Ser laws are fitted, can be used to deduce the structure of the central
profile at small radii.
Under this picture, excursions of the data above the \Ser model
are declared to be separate nuclear components, rather than as a simple failure
of the model.
The VCS nuclei effectively absorb the central flux left over from
the \Ser fits; the residual traces in VCS Figure 103
for such galaxies are nearly flat.
As Figure \ref{fig:acscomp3} demonstrates, forcing
a \Ser fit to galaxies that truly have slowly decreasing power-law profiles
will clearly create ersatz nuclei.
Again, we see no physical or even empirical justification to prefer the
\Ser $+$ nucleus model, let alone any demonstration strong enough
to show that the Nuker law-based interpretation is untenable.

The Nuker law, of course, has no physical justification either, however,
we are not using it to extrapolate beyond the domain of the fit;
it serves in this context as a smooth representation of the data.
Likewise, the use of \Ser models is interesting for the portions of the
envelope for which they are adequate fits.  Neither should be used
to extrapolate to radii considerably smaller than those that are well fitted
by their forms --- this is why we emphasize the use of $\gamma',$
which is always a local measure, in preference to $\gamma,$ which
itself is the slope only realized as $r\rightarrow0.$

In this context it is interesting to revisit the non-parametrically
derived space density profiles in Figure \ref{fig:den_pro}.  The
strong impression of bimodality in this figure is concordant with
the parametric analysis based on the limiting projected profile slopes
provided by the Nuker law.  Under the VCS interpretation, however,
the clean separation of the two bundles of profiles in the figure would
be interpreted as due to nuclear components in the upper bundle,
and the composite plot of ``true'' galaxy density profiles could only be
constructed with reference to the VCS parametric models.

In our analysis of WFPC1 data \citep{l95} or
WFPC2 data \citep{l05}, we identified nuclei by looking for
upturns above a power-law cusp as $r\rightarrow0.$
These are often obvious in core galaxies, as can be seen in some of the
examples in Figure \ref{fig:acscomp1}, but for power-law galaxies the
evidence for a possible nucleus is often just a subtle increase in the profile
slope as the resolution limit is approached.
There are no strong {\it upward} breaks in any of
the galaxies discussed in this section that would make detection
of a nucleus unambiguous.
At the same time, \citet{rest} present
deconvolved WFPC2 observations of NGC 4474 and 4482, two galaxies that
are plotted in Figure \ref{fig:gam_diff} with widely differing
$\gamma'$ values, that do have clearly defined nuclei,
but again ones more modest than those implied by the VCS models.

We explore the non-uniqueness of the
\Ser $+$ nucleus model in further detail for NGC 4458,
in which we plot
the \citet{l05} data for this galaxy extended to large radii by the
addition of the $B$ band data of \citet{jed} (Figure \ref{fig:n4458}).
The Nuker law is fitted to $r<15'',$ or 2.8 decades in radius,
a portion of the galaxy
that \citet{cote} consider to be a bright and extended nucleus.
The \citet{lf} \Ser fit in contrast is valid over only
$4''<r<50'',$ a little more than a single decade in radius, but is
then extrapolated inwards by 1.6 decades to the radius at which
\citet{lf} report $\gamma'.$

The profile of this object is obviously complex and the Nuker residuals
shown in \citet{l05} show a sinusoidal pattern, associated with an
increase in profile slope at $r\approx1''.$  Perhaps by our
criterion, this could be interpreted to be the onset of a nucleus.
Even so, the Nuker law fitted to just $r>1''$ would clearly model
this region with less curvature than implied by the \Ser model, again
giving a less luminous nucleus.  Yet another interpretation, however,
is that NGC 4458 is a dwarf S0 galaxy, in which the Nuker law
appropriately describes the bulge while the \Ser law correctly
captures the disk.
The case of NGC 4458 emphasizes that the use of \Ser models plus
large nuclei is neither unique nor compelling in galaxies of the luminosities
that we are considering.

\section{Summary}

The distribution of surface brightness cusp slopes as the {\it HST}
resolution limit is approached is strongly bimodal.  This confirms once again
the usefulness of classification of galaxies into core
or power-law systems based on their central brightness profiles.
Bimodality is still evident for galaxies within 50 Mpc, or for those
restricted in luminosity to $-22.5<M_V\leq-20.5.$  This last result
is of particular interest as it is the interval over which
the typical central structure of early-type galaxies changes
in form.  Very few core galaxies have $M_V\geq-20.5,$ while only a small
number of power-law galaxies have $M_V\leq-22.5.$  There are a few
galaxies that have profiles with $0.3<\gamma'<0.5,$
thus placing them in an intermediate position between core and power-law
systems.  Intriguingly, these galaxies are most common in the luminosity
range over which the core/power-law transition occurs.  Intermediate
galaxies are rare, however, and they do not fill in the ``valley''
in the distribution of cusp slopes.

As a check on our conclusions that the distribution
of cusp slopes is bimodal, we repeated the non-parametric analysis
of the {\it HST} brightness profiles presented by \citet{g96}.
This provides a strong demonstration of bimodality in the space
{\it density} profiles without reference to any fitting function or
any physical assumption beyond adopting the radius of maximum curvature
of the profiles as a common physical scale.  The concordance of this
very simple approach with the results from the Nuker law fits, shows
that inferences of bimodality from Nuker law fits alone are not due to
artifacts in or poor behavior of its parametric form,

These conclusions are inconsistent with the results of
\citet{lf}, who presented a strongly different distribution of $\gamma'$
values based on their VCS galaxies.
By examining galaxies in common with the VCS sample,
we find that profiles derived from ACS/WFC, WFPC2, and WFPC1 agree
extremely well to their stated resolution limits.
In contrast, we find that the VCS profile models, which the VCS uses
to characterize cusp slopes,
are either poor representations of the data
as the {\it HST} resolution limit is approached,
or consist of inward extrapolations of forms fitted to the envelope.
While the latter case, which requires the assumption of large nuclear
clusters, may be a plausible model, it is neither
unique nor compelling enough to demonstrate that the previous modeling
of profiles conducted by our group and others is ``untenable.''

We do note our large composite sample is largely
limited to galaxies with $M_V<-19.$
The core/power-law transition occurs at still larger luminosities,
and our conclusions on bimodality are limited to this context.
This luminosity limit corresponds to roughly
the median VCS galaxy luminosity, however, and many of the
faintest VCS galaxies have profiles that are nearly flat at the center.
\citet{lf} are concerned that under our criteria such systems would be
classified as cores, something that we have not actually done.
The envelope profiles of these galaxies are nearly exponential and have
no well-defined breaks as the center is approached\footnote{Indeed,
the maximum logarithmic second derivative, and hence the location
of $r_b,$ is reached only in the limit $r\rightarrow\infty$ for a
pure exponential profile.}.
Their overall forms
are completely different from the luminous core galaxies
and thus they are trivially distinguished from them.
\citet{k85}, \citet{k87}, and \citet{kb94} showed
that ellipticals and spheroidals form well defined, disjoint
sequences in parameter space.
A recent analysis of the profiles of 
low-luminosity ellipticals and a comparison to a large sample of Sph galaxies 
strongly confirms the E/Sph dichotomy \citep{k07}.
The resolution of this question, however,
is not relevant to the strong physical dichotomy between power-law
and core galaxies advanced in \citet{f97}.
Cores are a phenomenon of the most luminous galaxies.  Their likely immediate
progenitors are the power-law galaxies of slightly lower luminosity.
The distribution of $\gamma'$ for $M_V<-19$ is the relevant
diagnostic for how this transition occurred.

\acknowledgments

This research was supported in part by several grants provided through STScI
associated with GO programs 5512, 6099, 6587, 7388, 8591, 9106, and 9107.
Our team meetings were generously hosted
by the National Optical Astronomy Observatory, the Observatories
of the Carnegie Institution of Washington, the Aspen Center
for Physics, the Leiden Observatory, and the University of California, Santa
Cruz Center for Adaptive Optics.  We thank Steve Zepf for the KMM
software. We thank the referee for a careful review of the manuscript.
We thank Dr. Laura Ferrarese for her comments on an earlier draft of this paper.

\clearpage

\begin{deluxetable}{lcrccccl}
\tablecolumns{8}
\tablewidth{0pt}
\tablecaption{Global Parameters}
\tablehead{\colhead{ }&\colhead{ }
&\colhead{D}&\colhead{ }&\colhead{ }&\colhead{$R_e$}
&\colhead{ }&\colhead{ } \\
\colhead{Galaxy}&\colhead{Morph}
&\colhead{(Mpc)}&\colhead{Ref}&\colhead{$M_V$}&
\colhead{$\log({\rm pc})$}&\colhead{S}&\colhead{Notes}}
\startdata
NGC 0404        &S$0-$&  3.5&1&$-17.19$&3.17&6&$\Delta M=0.45$ (1)\\
NGC 0474        &S0& 29.3&2&$-20.12$&3.59&6&$\Delta M=0.82$ (2)\\
NGC 0507        &S0& 63.8&4&$-23.02$&4.11&1& \\
NGC 0524        &S$0+$& 25.4&1&$-21.85$&\dots&5&$\Delta M=0.19$ (2)\\
NGC 0545        &BCG& 73.3&2&$-22.98$&4.36&3&A0194-M1\\
NGC 0584        &E& 22.1&1&$-21.38$&3.53&1& \\
NGC 0596        &E& 22.1&1&$-20.90$&3.51&1& \\
NGC 0720        &E& 29.2&1&$-22.20$&3.80&4& \\
NGC 0741        &E& 70.9&4&$-23.27$&4.12&1& \\
NGC 0821        &E& 25.5&1&$-21.71$&3.56&1& \\
NGC 0910        &BCG& 81.1&2&$-22.79$&4.19&3&A0347-M1\\
NGC 1016        &E& 75.9&3&$-22.90$&\dots&1& \\
NGC 1023        &S$0-$& 12.1&1&$-20.53$&\dots&1&$\Delta M=0.73$ (3)\\
NGC 1052        &E& 20.5&1&$-21.17$&3.52&5& \\
NGC 1172        &E& 22.7&1&$-20.13$&3.64&4& \\
NGC 1316        &E& 22.7&1&$-23.32$&3.86&1& \\
NGC 1331        &E& 24.2&1&$-18.58$&\dots&5& \\
NGC 1351        &S$0-$& 21.0&1&$-20.08$&3.62&5& \\
NGC 1374        &E& 20.9&1&$-20.57$&3.41&1& \\
NGC 1399        &E& 21.1&1&$-22.07$&3.62&1& \\
NGC 1400        &S$0-$& 26.5&1&$-21.37$&3.47&4&$\Delta M=0.00$ (1)\\
NGC 1426        &E& 26.5&1&$-20.78$&3.56&1& \\
NGC 1427        &E& 21.0&1&$-20.79$&3.50&1& \\
NGC 1439        &E& 26.5&1&$-20.82$&3.61&1& \\
NGC 1500        &BCG&132.5&2&$-22.75$&4.19&3&A3193-M1\\
NGC 1553        &S0& 21.0&1&$-22.06$&\dots&5&$\Delta M=0.19$ (4)\\
NGC 1600        &E& 57.4&2&$-23.02$&4.13&4& \\
NGC 1700        &E& 40.6&2&$-21.95$&3.64&1& \\
NGC 2300        &S0& 30.4&4&$-21.74$&3.70&1& \\
NGC 2434        &E& 22.8&1&$-21.33$&3.56&1& \\
NGC 2549        &S0& 13.4&1&$-19.17$&3.37&2&$\Delta M=0.46$ (1)\\
NGC 2592        &E& 27.2&1&$-20.01$&\dots&2& \\
NGC 2634        &E& 35.3&1&$-20.83$&\dots&2& \\
NGC 2636        &E& 38.3&2&$-19.19$&\dots&4& \\
NGC 2685        &S$0+$& 14.3&2&$-19.72$&\dots&6&$\Delta M=0.00$ (1)\\
NGC 2699        &E& 28.4&1&$-20.25$&\dots&2& \\
NGC 2778        &E& 24.2&1&$-18.75$&3.21&1&$\Delta M=0.82$ (2)\\
NGC 2832        &BCG& 96.8&2&$-23.76$&4.50&3&A0779-M1\\
NGC 2841        &Sb& 15.1&2&$-20.57$&\dots&4&$\Delta M=1.54$ (5)\\
NGC 2872        &E& 47.4&2&$-21.62$&\dots&2& \\
NGC 2902        &S0& 33.0&2&$-20.59$&\dots&2&$\Delta M=0.00$ (1)\\
NGC 2907        &Sa& 34.4&4&$-21.23$&\dots&5&$\Delta M=0.00$ (1)\\
NGC 2950        &S0& 15.8&1&$-19.73$&3.19&2&$\Delta M=0.38$ (1)\\
NGC 2974        &E& 22.7&1&$-21.09$&3.61&1& \\
NGC 2986        &E& 37.6&2&$-22.32$&3.83&2& \\
NGC 3056        &S$0+$& 12.9&1&$-18.98$&\dots&5&$\Delta M=0.19$ (1)\\
NGC 3065        &S0& 29.6&2&$-19.64$&3.36&2&$\Delta M=0.35$ (1)\\
NGC 3078        &E& 37.2&1&$-21.95$&\dots&2& \\
NGC 3115        &S$0-$& 10.2&1&$-21.11$&3.37&1&$\Delta M=0.19$ (3)\\
NGC 3193        &E& 36.0&1&$-21.98$&3.63&2& \\
NGC 3266        &S0& 29.4&2&$-20.11$&\dots&2& \\
NGC 3348        &E& 41.5&2&$-22.18$&\dots&2& \\
NGC 3377        &E& 11.7&1&$-20.07$&3.30&1& \\
NGC 3379        &E& 11.7&1&$-21.14$&3.39&1& \\
NGC 3384        &S$0-$& 11.7&1&$-19.93$&3.08&1&$\Delta M=0.58$ (6)\\
NGC 3414        &S0& 26.7&1&$-20.25$&3.63&2&$\Delta M=0.97$ (1)\\
NGC 3551        &BCG&130.1&2&$-23.55$&4.79&3&A1177-M1\\
NGC 3585        &E& 21.2&1&$-21.93$&\dots&1& \\
NGC 3595        &E& 35.2&2&$-20.96$&\dots&2& \\
NGC 3599        &S0& 23.0&1&$-19.93$&\dots&4& \\
NGC 3605        &E& 23.0&1&$-19.61$&3.20&4& \\
NGC 3607        &S0& 10.9&1&$-19.88$&3.28&1&$\Delta M=0.48$ (1)\\
NGC 3608        &E& 23.0&1&$-21.12$&3.52&1& \\
NGC 3610        &E& 22.6&1&$-20.96$&3.16&1& \\
NGC 3613        &E& 30.8&1&$-21.59$&3.80&2& \\
NGC 3640        &E& 28.3&1&$-21.96$&3.69&1& \\
NGC 3706        &S$0-$& 46.9&4&$-22.31$&3.90&1& \\
NGC 3842        &BCG& 97.0&4&$-23.18$&4.25&1&A1367-M1\\
NGC 3900        &S$0+$& 29.9&2&$-20.80$&\dots&6&$\Delta M=0.33$ (1)\\
NGC 3945        &S$0+$& 19.9&3&$-20.25$&\dots&1&$\Delta M=0.49$ (1)\\
NGC 4026        &S0& 15.6&1&$-19.79$&3.59&1&$\Delta M=0.50$ (1)\\
NGC 4073        &E& 92.2&4&$-23.50$&\dots&1& \\
NGC 4121        &E& 25.2&1&$-18.53$&\dots&2& \\
NGC 4128        &S0& 34.5&2&$-20.79$&3.48&2& \\
NGC 4143        &S0& 15.6&1&$-19.68$&3.13&6&$\Delta M=0.59$ (1)\\
NGC 4150        &S0& 14.5&1&$-18.66$&\dots&5&$\Delta M=0.57$ (1)\\
NGC 4168        &E& 37.3&2&$-21.80$&3.92&2& \\
NGC 4239        &E& 17.5&2&$-18.50$&\dots&4& \\
NGC 4261        &E& 33.4&1&$-22.26$&3.91&5& \\
NGC 4278        &E& 16.7&1&$-21.05$&3.39&1& \\
NGC 4291        &E& 25.0&1&$-20.64$&3.29&1& \\
NGC 4365        &E& 21.6&1&$-22.18$&3.89&1& \\
NGC 4374        &E& 17.9&1&$-22.28$&3.75&5& \\
NGC 4382        &S$0+$& 17.9&1&$-21.96$&3.81&1&$\Delta M=0.29$ (1)\\
NGC 4387        &E& 17.9&1&$-19.25$&3.23&4& \\
NGC 4406        &E& 17.9&1&$-22.46$&4.09&1& \\
NGC 4417        &S0& 17.9&1&$-18.94$&3.50&6&$\Delta M=1.30$ (1)\\
NGC 4434        &E& 17.9&1&$-19.19$&\dots&4& \\
NGC 4458        &E& 17.9&1&$-19.27$&\dots&1& \\
NGC 4464        &E& 17.9&1&$-18.82$&\dots&4& \\
NGC 4467        &E& 17.9&1&$-17.51$&\dots&4& \\
NGC 4472        &E& 17.9&1&$-22.93$&3.94&1& \\
NGC 4473        &E& 17.9&1&$-21.16$&3.42&1& \\
NGC 4474        &S0& 21.6&1&$-18.42$&3.30&2&$\Delta M=1.89$ (1)\\
NGC 4478        &E& 17.9&1&$-19.89$&3.13&1& \\
NGC 4482        &E& 17.9&1&$-18.87$&\dots&2& \\
NGC 4486        &E& 17.9&1&$-22.71$&3.94&4& \\
NGC 4486B       &cE& 17.9&1&$-17.98$&\dots&1& \\
NGC 4494        &E& 17.9&1&$-21.50$&3.57&1& \\
NGC 4503        &S$0-$& 17.9&1&$-19.57$&\dots&2&$\Delta M=0.79$ (1)\\
NGC 4551        &E& 17.9&1&$-19.37$&3.32&4& \\
NGC 4552        &E& 17.9&1&$-21.65$&3.55&1& \\
NGC 4564        &E& 17.9&1&$-20.26$&3.44&2& \\
NGC 4589        &E& 25.0&1&$-21.35$&3.71&1& \\
NGC 4621        &E& 17.9&1&$-21.74$&3.57&1& \\
NGC 4636        &E& 17.9&1&$-21.86$&4.02&5& \\
NGC 4648        &E& 27.7&1&$-20.24$&\dots&2& \\
NGC 4649        &E& 17.9&1&$-22.51$&3.82&1& \\
NGC 4660        &E& 17.9&1&$-20.13$&3.02&1& \\
NGC 4696        &BCG& 40.0&2&$-24.33$&4.52&3&A3526-M1\\
NGC 4697        &E& 13.3&1&$-21.49$&3.71&4& \\
NGC 4709        &E& 37.0&1&$-22.32$&\dots&1& \\
NGC 4742        &E& 16.4&1&$-19.90$&2.80&4& \\
NGC 4874        &E&106.6&2&$-23.49$&4.79&4& \\
NGC 4889        &BCG& 95.0&2&$-23.73$&4.40&3&A1656-M1\\
NGC 5017        &E& 40.3&2&$-20.67$&\dots&2& \\
NGC 5061        &E& 26.9&4&$-22.01$&3.55&1& \\
NGC 5077        &E& 44.9&2&$-22.07$&\dots&2& \\
NGC 5198        &E& 38.3&2&$-21.23$&\dots&2& \\
NGC 5308        &S$0-$& 33.0&1&$-21.26$&3.90&2& \\
NGC 5370        &S0& 45.2&2&$-20.60$&\dots&2& \\
NGC 5419        &E& 62.6&4&$-23.37$&4.18&1& \\
NGC 5557        &E& 52.7&4&$-22.62$&3.77&1& \\
NGC 5576        &E& 27.1&1&$-21.31$&3.58&1& \\
NGC 5796        &E& 45.2&2&$-21.98$&\dots&2& \\
NGC 5812        &E& 28.4&1&$-21.39$&\dots&2& \\
NGC 5813        &E& 28.7&1&$-22.01$&3.94&1& \\
NGC 5831        &E& 28.7&1&$-21.00$&3.53&2& \\
NGC 5838        &S$0-$& 22.2&2&$-20.51$&3.83&6&$\Delta M=0.45$ (1)\\
NGC 5845        &E& 28.7&1&$-19.98$&2.66&5& \\
NGC 5898        &E& 33.3&1&$-21.65$&\dots&2& \\
NGC 5903        &E& 33.3&1&$-21.90$&\dots&2& \\
NGC 5982        &E& 40.5&4&$-21.97$&3.70&1& \\
NGC 6086        &BCG&131.2&2&$-23.11$&4.37&3&A2162-M1\\
NGC 6166        &BCG&128.6&2&$-23.80$&4.83&4&A2199-M1\\
NGC 6173        &BCG&126.2&2&$-23.59$&4.47&3&A2197-M1\\
NGC 6278        &S0& 40.0&2&$-20.81$&\dots&2& \\
NGC 6340        &S0& 16.8&2&$-19.46$&\dots&6&$\Delta M=0.82$ (1)\\
NGC 6849        &S0& 84.0&4&$-22.78$&4.07&1& \\
NGC 6876        &E& 57.6&4&$-23.58$&\dots&1& \\
NGC 7014        &BCG& 71.1&2&$-22.18$&3.83&3&A3742-M1\\
NGC 7052        &E& 67.1&2&$-22.35$&4.12&5& \\
NGC 7213        &Sa& 22.7&3&$-21.71$&\dots&1& \\
NGC 7332        &S0& 24.3&1&$-19.62$&3.20&4&$\Delta M=1.32$ (1)\\
NGC 7457        &S$0-$& 14.0&1&$-18.62$&3.83&1&$\Delta M=1.08$ (3)\\
NGC 7578B       &BCG&171.9&2&$-23.41$&4.57&3&A2572-M1\\
NGC 7619        &E& 56.1&1&$-22.94$&3.97&1& \\
NGC 7626        &E& 56.0&1&$-22.87$&4.08&5& \\
NGC 7647        &BCG&171.4&2&$-23.97$&\dots&3&A2589-M1\\
NGC 7727        &Sa& 21.6&3&$-21.19$&\dots&1& \\
NGC 7743        &S$0+$& 21.9&1&$-20.18$&\dots&6&$\Delta M=0.27$ (1)\\
NGC 7785        &E& 49.9&4&$-22.08$&3.62&1& \\
IC 0115 &BCG&171.7&2&$-22.67$&4.19&3&A0195-M1\\
IC 0613 &BCG&128.9&2&$-22.27$&4.32&3&A1016-M1\\
IC 0664 &BCG&138.9&2&$-22.86$&4.37&3&A1142-M1\\
IC 0712 &BCG&140.7&2&$-23.29$&4.85&3&A1314-M1\\
IC 0875 &S0& 41.9&2&$-20.21$&\dots&2& \\
IC 1459 &E& 30.9&1&$-22.51$&3.65&1& \\
IC 1565 &BCG& 38.2&2&$-22.99$&3.75&3&A0076-M1\\
IC 1633 &BCG& 98.4&2&$-23.91$&4.31&3&A2877-M1\\
IC 1695 &BCG&192.0&2&$-23.90$&4.74&3&A0193-M1\\
IC 1733 &BCG&150.4&2&$-23.43$&4.59&3&A0260-M1\\
IC 2738 &BCG&152.7&2&$-22.18$&4.26&3&A1228-M1\\
IC 4329 &BCG& 56.7&4&$-23.95$&4.62&1&A3574-M1\\
IC 4931 &BCG& 83.3&2&$-23.47$&4.53&3&A3656-M1\\
IC 5353 &BCG&124.0&2&$-22.64$&4.18&3&A4038-M1\\
A0119-M1&BCG&177.9&2&$-24.01$&4.66&3& \\
A0168-M1&BCG&177.9&2&$-23.12$&4.50&3& \\
A0189-M1&BCG&133.4&2&$-21.89$&3.97&3& \\
A0261-M1&BCG&181.6&2&$-22.95$&4.36&3& \\
A0295-M1&BCG&168.1&2&$-23.11$&4.60&3& \\
A0376-M1&BCG&195.2&2&$-23.60$&4.68&3& \\
A0397-M1&BCG&132.9&2&$-23.42$&4.53&3& \\
A0419-M1&BCG&156.0&2&$-21.79$&3.91&3& \\
A0496-M1&BCG&126.9&2&$-24.28$&4.77&3& \\
A0533-M1&BCG&179.4&2&$-22.68$&4.33&3& \\
A0548-M1&BCG&152.6&2&$-22.75$&4.32&3& \\
A0634-M1&BCG&117.3&2&$-22.70$&4.22&3& \\
A0912-M1&BCG&176.0&2&$-22.24$&3.98&3& \\
A0999-M1&BCG&128.9&2&$-22.45$&4.14&3& \\
A1020-M1&BCG&278.6&2&$-22.65$&\dots&4& \\
A1631-M1&BCG&176.7&2&$-23.34$&4.51&3& \\
A1831-M1&BCG&321.0&2&$-23.51$&\dots&4& \\
A1983-M1&BCG&176.1&2&$-22.35$&3.87&3& \\
A2040-M1&BCG&176.6&2&$-23.46$&4.71&3& \\
A2052-M1&BCG&150.9&2&$-23.04$&5.02&4& \\
A2147-M1&BCG&140.0&2&$-23.16$&4.56&3& \\
A2247-M1&BCG&163.3&2&$-22.66$&4.17&3& \\
A3144-M1&BCG&173.0&2&$-22.28$&4.04&3& \\
A3376-M1&BCG&176.0&2&$-23.29$&4.55&3& \\
A3395-M1&BCG&186.1&2&$-24.23$&4.91&3& \\
A3528-M1&BCG&204.6&2&$-24.30$&5.02&3& \\
A3532-M1&BCG&208.6&2&$-24.58$&5.06&3& \\
A3554-M1&BCG&180.9&2&$-23.99$&4.72&3& \\
A3556-M1&BCG&183.0&2&$-23.65$&4.30&3& \\
A3558-M1&BCG&180.8&2&$-24.92$&4.91&3& \\
A3562-M1&BCG&185.6&2&$-24.32$&5.11&3& \\
A3564-M1&BCG&186.2&2&$-22.68$&3.75&3& \\
A3570-M1&BCG&142.3&2&$-22.54$&4.37&3& \\
A3571-M1&BCG&151.3&2&$-25.30$&5.33&3& \\
A3677-M1&BCG&186.6&2&$-22.21$&4.42&3& \\
A3716-M1&BCG&180.2&2&$-23.75$&4.73&3& \\
A3736-M1&BCG&196.2&2&$-23.98$&4.62&3& \\
A3747-M1&BCG&128.6&2&$-22.65$&4.16&3& \\
ESO 378-20      &S0& 47.7&2&$-20.97$&\dots&2& \\
ESO 443-39      &S0& 47.2&2&$-20.93$&\dots&2& \\
ESO 447-30      &S0& 41.7&2&$-21.17$&\dots&2& \\
ESO 462-15      &E& 80.3&4&$-22.83$&3.86&1& \\
ESO 507-27      &S0& 50.1&2&$-20.89$&\dots&2& \\
ESO 507-45      &S0& 73.3&4&$-23.28$&\dots&5& \\
MCG 11-14-25A   &E& 49.0&2&$-19.08$&\dots&2& \\
MCG 8-27-18     &S0& 48.7&2&$-20.03$&\dots&2& \\
UGC 4551        &S0& 27.1&2&$-19.78$&\dots&2& \\
UGC 4587        &S0& 45.9&2&$-20.77$&\dots&2& \\
UGC 6062        &S0& 42.2&2&$-20.34$&\dots&2& \\
VCC 1199        &E& 17.9&1&$-15.58$&\dots&4& \\
VCC 1440        &E& 17.9&1&$-17.24$&\dots&4& \\
VCC 1545        &E& 17.9&1&$-17.49$&\dots&4& \\
VCC 1627        &E& 17.9&1&$-16.42$&\dots&4& \\
\enddata
\label{tab:glob}
\tablecomments{Morphological classifications are from the RC3 \citep{rc3}.
Distance reference codes: 1) \citet{ton} SBF distances
adjusted to $H_0=70$ km s$^{-1}$ Mpc$^{-1}$; 2) distance from
\citet{f97} scaled to $H_0=70;$ 3) \citet{f89} group distance;
and 4) distance based on CMB velocity of galaxy or group.
Galaxy absolute luminosities are based on RC3 $V_T$ values, or
the \citet{laine} $R$ magnitudes for the BGC; \citet{sfd} extinction
corrections have been applied.
The (S) column refers to the source of central structural parameters
for the given galaxy as follows: 1) \citet{l05}; 2) \citet{rest}:
3) \citet{laine}; 4) \citet{l95}; 5) \citet{quil}; and 6) \citet{rav}.
$\Delta M$ values in the notes column refer to the luminosity corrections
applied to subtract the disks in S0 galaxies; reference codes in the
parentheses correspond to 1) \cite{bag}, 2) \citet{bag}, 3) \citet{kent},
4) \citet{k83}, 5) \citet{bor}, and 6) \citet{bur}.}
\end{deluxetable}

\begin{deluxetable}{lccrrcccrrl}
\tablecolumns{11}
\tablewidth{0pt}
\tablecaption{Central Structural Parameters}
\tablehead{\colhead{ }&\colhead{ }&\colhead{ }&\colhead{$r_b$}
&\colhead{$\theta_b$}&\colhead{ }&\colhead{ }
&\colhead{ }&\colhead{ }&\colhead{ }&\colhead{ } \\
\colhead{Galaxy}&\colhead{S}&\colhead{P}&\colhead{(pc)}
&\colhead{(arcsec)}&\colhead{$I_b$}&\colhead{$\alpha$}
&\colhead{$\beta$}&\colhead{$\gamma$}&\colhead{$\gamma'$}&\colhead{Notes}}
\startdata
NGC 0404       &6&$\backslash$&   1.7&0.40&16.01&0.0&1.58& 0.28& 0.92&$\cap$ in ref. (6)\\
NGC 0474       &6&$\backslash$&  14.2&1.47&17.17&1.2&1.90& 0.37& 0.56&$\wedge$ in ref. (6)\\
NGC 0507       &1&$\cap$& 343.3&1.11&17.16&1.3&1.75& 0.00& 0.01& \\
NGC 0524       &6&$\cap$& 173.6&1.41&16.58&0.7&1.69& 0.03& 0.27& \\
NGC 0545       &3&$\cap$& 248.8&0.70&16.90&1.5&1.47& 0.08& 0.10& \\
NGC 0584       &1&$\cap$&  47.1&0.44&15.22&0.5&1.61&$-$0.01& 0.30& \\
NGC 0596       &1&$\backslash$&   4.3&0.54&15.90&0.5&1.59& 0.16& 0.54& \\
NGC 0720       &4&$\cap$& 487.0&3.44&17.45&2.2&1.43& 0.06& 0.07& \\
NGC 0741       &1&$\cap$& 391.9&1.14&17.67&2.3&1.29& 0.10& 0.11& \\
NGC 0821       &1&$\wedge$&  59.3&0.48&15.69&0.4&1.71& 0.10& 0.42&$\backslash$ in ref. (6)\\
NGC 0910       &3&$\cap$& 153.3&0.39&16.93&4.2&0.91&$-$0.03&$-$0.03& \\
NGC 1016       &1&$\cap$& 618.2&1.68&18.01&1.0&1.90& 0.09& 0.11& \\
NGC 1023       &1&$\backslash$&   2.3&1.23&15.71&7.2&1.15& 0.74& 0.74& \\
NGC 1052       &5&$\cap$&  43.7&0.44&14.62&2.2&1.27& 0.18& 0.22& \\
NGC 1172       &4&$\backslash$&   4.4&0.54&16.66&0.3&2.21&$-$0.01& 0.86& \\
NGC 1316       &1&$\wedge$&  79.2&0.72&14.86&2.3&1.14& 0.35& 0.35&$\cap$ in ref. (4)\\
NGC 1331       &5&$\backslash$&  11.7&2.91&19.32&3.4&1.29& 0.57& 0.57& \\
NGC 1351       &5&$\backslash$&  10.2&1.00&16.26&3.8&1.31& 0.78& 0.78& \\
NGC 1374       &1&$\cap$&   9.1&0.09&14.57&1.9&1.08&$-$0.09&$-$0.03& \\
NGC 1399       &1&$\cap$& 239.4&2.34&16.96&2.3&1.32& 0.12& 0.12& \\
NGC 1400       &4&$\cap$&  39.8&0.31&15.43&1.2&1.35&$-$0.10& 0.20&$\backslash$ in ref. (5)\\
NGC 1426       &1&$\backslash$&   5.1&2.20&17.62&0.5&1.96& 0.26& 0.57& \\
NGC 1427       &1&$\backslash$&   4.1&0.86&16.25&0.8&1.67& 0.30& 0.51& \\
NGC 1439       &1&$\backslash$&   5.1&0.65&16.21&4.8&1.48& 0.74& 0.74& \\
NGC 1500       &3&$\cap$& 167.0&0.26&16.70&1.4&1.37& 0.08& 0.20& \\
NGC 1553       &5&$\backslash$&  10.2&5.72&16.93&3.8&1.43& 0.74& 0.74& \\
NGC 1600       &4&$\cap$&1870.0&6.72&19.03&1.1&2.12&$-$0.03& 0.00& \\
NGC 1700       &1&$\cap$&  11.8&0.06&13.58&1.0&1.19&$-$0.10& 0.07&$\backslash$ in ref. (4)\\
NGC 2300       &1&$\cap$& 330.1&2.24&17.53&1.2&1.80& 0.07& 0.08& \\
NGC 2434       &1&$\backslash$&   4.4&5.41&19.10&1.9&2.05& 0.75& 0.75& \\
NGC 2549       &2&$\backslash$&   3.2&3.70&17.57&1.8&1.71& 0.67& 0.67& \\
NGC 2592       &2&$\backslash$&   6.6&1.37&17.22&3.3&1.60& 0.92& 0.92& \\
NGC 2634       &2&$\backslash$&   8.6&1.82&17.93&2.8&1.57& 0.81& 0.81& \\
NGC 2636       &4&$\backslash$&   7.4&0.24&16.53&0.6&1.66& 0.06& 0.67& \\
NGC 2685       &6&$\backslash$&   6.9&2.38&17.03&1.7&1.52& 0.73& 0.73& \\
NGC 2699       &2&$\backslash$&   6.9&2.66&18.24&1.7&1.89& 0.84& 0.85& \\
NGC 2778       &1&$\backslash$&   4.7&0.35&16.06&0.4&1.75& 0.33& 0.83& \\
NGC 2832       &3&$\cap$& 488.1&1.04&17.37&1.7&1.44& 0.02& 0.03& \\
NGC 2841       &4&$\wedge$&  11.7&0.16&14.52&1.6&0.92& 0.05& 0.34&$\backslash$ in ref. (4)\\
NGC 2872       &2&$\backslash$&  11.5&4.27&18.72&2.5&1.66& 1.01& 1.01& \\
NGC 2902       &2&$\wedge$& 140.8&0.88&16.95&1.9&1.57& 0.49& 0.50&$\backslash$ in ref. (2) \\
NGC 2907       &5&$\backslash$&  16.7&1.96&16.60&0.5&1.78& 0.58& 0.79& \\
NGC 2950       &2&$\backslash$&   3.8&2.43&16.77&2.4&1.81& 0.82& 0.82& \\
NGC 2974       &1&$\backslash$&   4.4&0.31&15.17&25.1&1.05& 0.62& 0.62& \\
NGC 2986       &2&$\cap$& 226.0&1.24&16.70&1.8&1.50& 0.18& 0.18& \\
NGC 3056       &5&$\backslash$&   6.3&4.11&18.05&2.1&1.80& 0.90& 0.90& \\
NGC 3065       &2&$\backslash$&   7.2&3.14&18.58&1.0&2.28& 0.79& 0.84& \\
NGC 3078       &2&$\backslash$&   9.0&3.43&17.80&2.4&1.60& 0.95& 0.95& \\
NGC 3115       &1&$\backslash$&   2.0&0.86&14.48&4.3&1.09& 0.52& 0.52& \\
NGC 3193       &2&$\cap$& 141.4&0.81&16.07&0.6&1.89& 0.01& 0.28&$\wedge$ in ref. (2) \\
NGC 3266       &2&$\backslash$&   7.1&2.25&18.39&1.3&2.06& 0.64& 0.66& \\
NGC 3348       &2&$\cap$& 199.2&0.99&16.61&1.2&1.53& 0.09& 0.12& \\
NGC 3377       &1&$\backslash$&   2.3&0.31&14.12&0.3&1.96& 0.03& 0.62& \\
NGC 3379       &1&$\cap$& 112.3&1.98&16.15&1.5&1.54& 0.18& 0.18& \\
NGC 3384       &1&$\backslash$&   2.3&3.15&16.44&15.3&1.81& 0.71& 0.71& \\
NGC 3414       &2&$\backslash$&   6.5&1.72&17.07&1.4&1.45& 0.83& 0.84& \\
NGC 3551       &3&$\cap$& 359.5&0.57&17.64&1.8&1.31& 0.13& 0.14& \\
NGC 3585       &1&$\wedge$&  37.0&0.36&14.72&1.6&1.06& 0.31& 0.31& \\
NGC 3595       &2&$\backslash$&   8.5&2.30&18.04&2.3&1.52& 0.75& 0.76& \\
NGC 3599       &4&$\backslash$&   4.5&1.25&17.45&50.0&1.45& 0.75& 0.75& \\
NGC 3605       &4&$\backslash$&   4.5&0.97&17.21&2.4&1.27& 0.59& 0.60& \\
NGC 3607       &1&$\cap$& 128.4&2.43&16.87&2.1&1.70& 0.26& 0.26& \\
NGC 3608       &1&$\cap$&  53.5&0.48&15.73&0.9&1.50& 0.09& 0.17& \\
NGC 3610       &1&$\backslash$&   4.4&2.84&16.39&48.5&1.86& 0.76& 0.76& \\
NGC 3613       &2&$\cap$&  50.8&0.34&15.72&1.5&1.06& 0.04& 0.08& \\
NGC 3640       &1&$\cap$&  46.6&0.34&15.66&0.6&1.30&$-$0.10& 0.03&$\wedge$ in ref. (2)\\
NGC 3706       &1&$\cap$&  40.9&0.18&14.18&14.7&1.24&$-$0.01&$-$0.01& \\
NGC 3842       &3&$\cap$& 399.9&0.93&17.60&2.2&1.27& 0.11& 0.12& \\
NGC 3900       &6&$\backslash$&  14.5&0.23&15.20&0.3&1.66& 0.51& 1.02& \\
NGC 3945       &1&$\backslash$&   3.9&7.38&18.62&0.3&2.56&$-$0.06& 0.57& \\
NGC 4026       &1&$\backslash$&   3.0&0.63&15.23&0.4&1.78& 0.15& 0.65& \\
NGC 4073       &1&$\cap$& 129.6&0.29&16.53&4.4&0.99&$-$0.08&$-$0.08& \\
NGC 4121       &2&$\backslash$&   6.1&5.60&20.30&1.5&3.65& 0.85& 0.85& \\
NGC 4128       &2&$\backslash$&   8.4&1.59&16.92&1.1&1.69& 0.71& 0.75& \\
NGC 4143       &6&$\backslash$&   7.6&3.11&17.11&1.3&2.18& 0.59& 0.61& \\
NGC 4150       &6&$\backslash$&   8.4&0.63&15.80&1.2&1.67& 0.58& 0.68& \\
NGC 4168       &2&$\cap$& 365.3&2.02&18.06&1.4&1.39& 0.17& 0.17& \\
NGC 4239       &2&$\wedge$&29.7&0.35&17.38&0.3&1.25&-0.06& 0.46& \\
NGC 4261       &6&$\cap$& 262.3&1.62&16.43&2.4&1.43& 0.00& 0.00& \\
NGC 4278       &1&$\cap$& 102.0&1.26&16.20&1.4&1.46& 0.06& 0.10& \\
NGC 4291       &1&$\cap$&  72.7&0.60&15.66&1.5&1.60& 0.01& 0.02& \\
NGC 4365       &1&$\cap$& 231.4&2.21&16.88&1.7&1.52& 0.07& 0.09& \\
NGC 4374       &6&$\cap$& 207.4&2.39&16.20&2.1&1.50& 0.13& 0.13& \\
NGC 4382       &1&$\cap$&  80.7&0.93&15.67&1.1&1.39& 0.00& 0.01& \\
NGC 4387       &4&$\backslash$&   3.5&0.44&16.85&0.2&1.37& 0.10& 0.65& \\
NGC 4406       &1&$\cap$&  79.8&0.92&16.03&3.9&1.04&$-$0.04&$-$0.04& \\
NGC 4417       &6&$\backslash$&   8.7&3.66&17.59&0.9&1.77& 0.71& 0.75& \\
NGC 4434       &4&$\backslash$&   3.5&0.61&16.57&0.4&1.97&$-$0.04& 0.64& \\
NGC 4458       &1&$\cap$&   7.8&0.09&13.72&4.5&1.40& 0.16& 0.17&$\backslash$ in ref. (4)\\
NGC 4464       &4&$\backslash$&   3.5&0.60&16.23&0.5&1.99& 0.14& 0.70& \\
NGC 4467       &4&$\backslash$&   3.5&10.72&22.91&2.2&7.14& 0.94& 0.94& \\
NGC 4472       &1&$\cap$& 211.7&2.44&16.63&1.9&1.17& 0.01& 0.01& \\
NGC 4473       &1&$\cap$& 386.2&4.45&17.24&0.7&2.60&$-$0.07& 0.01& \\
NGC 4474       &2&$\backslash$&   5.2&4.55&18.58&2.3&1.65& 0.72& 0.72& \\
NGC 4478       &1&$\cap$&  19.1&0.22&15.46&1.9&1.01&$-$0.10& 0.10&$\backslash$ in ref. (4)\\
NGC 4482       &2&$\wedge$& 350.6&4.04&20.24&3.4&1.01& 0.49& 0.49&$\backslash$ in ref. (2) \\
NGC 4486       &4&$\cap$& 491.2&5.66&17.32&8.4&1.06& 0.27& 0.27& \\
NGC 4486B      &1&$\cap$&  13.9&0.16&14.53&2.8&1.39&$-$0.10&$-$0.10& \\
NGC 4494       &1&$\backslash$&   3.5&2.82&17.19&0.7&1.88& 0.52& 0.55& \\
NGC 4503       &2&$\backslash$&   4.3&1.65&17.14&1.8&1.30& 0.64& 0.65& \\
NGC 4551       &4&$\backslash$&   3.5&0.34&16.49&0.2&1.64&$-$0.07& 0.69& \\
NGC 4552       &1&$\cap$&  42.5&0.49&15.20&2.0&1.17&$-$0.10&$-$0.02& \\
NGC 4564       &2&$\backslash$&   4.3&1.28&16.50&1.4&1.27& 0.80& 0.81& \\
NGC 4589       &1&$\cap$&  67.9&0.56&16.08&1.1&1.32& 0.21& 0.25&$\wedge$ in ref. (2) \\
NGC 4621       &1&$\backslash$&   3.5&1.00&15.76&0.4&1.34& 0.75& 0.85& \\
NGC 4636       &6&$\cap$& 298.5&3.44&17.45&1.7&1.56& 0.13& 0.13& \\
NGC 4648       &2&$\backslash$&   6.7&1.02&16.48&3.7&1.54& 0.92& 0.92& \\
NGC 4649       &1&$\cap$& 388.8&4.48&17.17&1.9&1.46& 0.17& 0.17& \\
NGC 4660       &1&$\backslash$&   3.5&1.76&16.34&5.6&1.50& 0.91& 0.91& \\
NGC 4696       &3&$\cap$& 271.5&1.40&17.76&6.6&0.86& 0.10& 0.10& \\
NGC 4697       &4&$\backslash$&   2.6&1.49&16.34&0.4&1.25& 0.22& 0.51& \\
NGC 4709       &1&$\wedge$& 227.8&1.27&17.48&2.2&1.43& 0.32& 0.32& \\
NGC 4742       &4&$\backslash$&   3.2&2.00&16.86&50.0&2.13& 1.04& 1.04& \\
NGC 4874       &4&$\cap$&1731.3&3.35&19.40&1.9&1.61& 0.12& 0.12& \\
NGC 4889       &3&$\cap$& 967.2&2.10&18.03&2.1&1.46& 0.03& 0.03& \\
NGC 5017       &2&$\backslash$&   9.8&1.48&17.53&3.0&1.59& 1.12& 1.12& \\
NGC 5061       &1&$\cap$&  37.8&0.29&14.35&1.6&1.43& 0.04& 0.05& \\
NGC 5077       &2&$\cap$& 350.5&1.61&17.10&1.1&1.67& 0.23& 0.26&$\wedge$ in ref. (2) \\
NGC 5198       &2&$\cap$&  29.7&0.16&15.40&2.6&1.13& 0.23& 0.26&$\wedge$ in ref. (2)\\
NGC 5308       &2&$\backslash$&   8.0&0.83&16.16&0.4&1.27& 0.82& 0.96& \\
NGC 5370       &2&$\backslash$&  11.0&1.63&18.45&0.8&1.50& 0.62& 0.67& \\
NGC 5419       &1&$\cap$& 722.3&2.38&17.87&1.4&1.69&$-$0.10& 0.03& \\
NGC 5557       &1&$\cap$& 204.4&0.80&16.40&0.8&1.68& 0.02& 0.07&$\wedge$ in ref. (2)\\
NGC 5576       &1&$\cap$& 549.2&4.18&17.81&0.4&2.73& 0.01& 0.26&$\backslash$ in ref. (2)\\
NGC 5796       &2&$\wedge$& 232.3&1.06&16.80&0.8&1.67& 0.41& 0.49&$\backslash$ in ref. (2)\\
NGC 5812       &2&$\backslash$&   6.9&1.84&17.22&1.2&1.67& 0.59& 0.62& \\
NGC 5813       &1&$\cap$& 160.0&1.15&16.82&1.6&1.60& 0.05& 0.06& \\
NGC 5831       &2&$\backslash$&   7.0&1.78&17.51&0.5&1.84& 0.33& 0.55& \\
NGC 5838       &6&$\backslash$&  10.8&4.35&17.69&2.6&1.87& 0.93& 0.93& \\
NGC 5845       &5&$\backslash$&  13.9&1.38&15.86&2.1&2.18& 0.51& 0.52& \\
NGC 5898       &2&$\wedge$& 200.2&1.24&16.93&1.2&1.57& 0.41& 0.43&$\backslash$ in ref. (2)\\
NGC 5903       &2&$\cap$& 256.7&1.59&17.45&1.8&1.48& 0.13& 0.13& \\
NGC 5982       &1&$\cap$&  68.7&0.35&15.66&2.1&1.03& 0.05& 0.05& \\
NGC 6086       &3&$\cap$& 661.5&1.04&17.75&1.4&1.78&$-$0.01& 0.02& \\
NGC 6166       &4&$\cap$&1502.6&2.41&19.33&4.3&0.91& 0.12& 0.12& \\
NGC 6173       &3&$\cap$& 477.2&0.78&17.28&0.6&1.86&$-$0.33& 0.02& \\
NGC 6278       &2&$\backslash$&   9.7&0.60&16.30&0.8&1.62& 0.55& 0.67& \\
NGC 6340       &6&$\backslash$&   8.1&0.28&15.39&2.5&1.28& 0.59& 0.64& \\
NGC 6849       &1&$\cap$& 138.5&0.34&17.03&1.0&1.21& 0.01& 0.08& \\
NGC 6876       &1&$\cap$& 139.6&0.50&16.99&12.1&0.75& 0.00& 0.00& \\
NGC 7014       &3&$\cap$&  75.8&0.22&15.61&2.2&1.05& 0.08& 0.11& \\
NGC 7052       &5&$\cap$& 250.5&0.77&16.35&2.8&1.19& 0.16& 0.16& \\
NGC 7213       &1&$\cap$& 383.0&3.48&17.71&1.0&3.10& 0.06& 0.21& \\
NGC 7332       &4&$\backslash$&   4.7&1.01&15.91&0.6&1.59& 0.62& 0.80& \\
NGC 7457       &1&$\backslash$&   2.7&0.22&16.33&1.0&1.05&$-$0.10& 0.61&$\wedge$ in ref. (6)\\
NGC 7578B      &3&$\cap$& 150.0&0.18&16.35&1.7&1.12& 0.10& 0.21& \\
NGC 7619       &1&$\cap$& 223.0&0.82&16.41&1.1&1.63&$-$0.02& 0.01& \\
NGC 7626       &6&$\wedge$& 160.2&0.59&16.09&1.8&1.30& 0.36& 0.40&$\backslash$ in ref. (5)\\
NGC 7647       &3&$\cap$& 207.7&0.25&17.19&3.4&1.10& 0.04& 0.05& \\
NGC 7727       &1&$\wedge$& 601.1&5.74&18.42&0.6&1.93& 0.43& 0.49& \\
NGC 7743       &6&$\backslash$&  10.6&0.16&14.44&5.4&1.38& 0.50& 0.57& \\
NGC 7785       &1&$\cap$&  16.9&0.07&15.14&0.9&0.97&$-$0.10& 0.06& \\
IC 0115        &3&$\cap$& 407.9&0.49&17.49&2.1&1.41& 0.09& 0.10& \\
IC 0613        &3&$\cap$& 143.7&0.23&16.41&3.9&1.16& 0.24& 0.25& \\
IC 0664        &3&$\cap$& 127.9&0.19&15.86&9.3&1.31& 0.12& 0.12& \\
IC 0712        &3&$\cap$& 811.7&1.19&18.06&2.3&1.60& 0.17& 0.17& \\
IC 0875        &2&$\backslash$&  10.2&1.87&18.26&0.8&1.81& 1.07& 1.12& \\
IC 1459        &1&$\cap$& 293.6&1.96&16.36&0.8&2.08&$-$0.10& 0.15& \\
IC 1565        &3&$\cap$&  51.9&0.28&16.94&1.2&1.33&$-$0.20&$-$0.03& \\
IC 1633        &3&$\cap$& 467.5&0.98&17.04&1.3&1.58&$-$0.02& 0.01& \\
IC 1695        &3&$\cap$& 363.0&0.39&17.01&2.8&1.48& 0.23& 0.23& \\
IC 1733        &3&$\cap$& 546.9&0.75&17.71&3.4&1.29&$-$0.01&$-$0.01& \\
IC 2738        &3&$\backslash$&  37.0&0.08&16.61&1.4&1.41& 0.53& 0.60& \\
IC 4329        &3&$\cap$& 208.9&0.76&16.97&2.5&1.30&$-$0.02&$-$0.02& \\
IC 4931        &3&$\cap$& 399.8&0.99&17.14&2.1&1.46& 0.09& 0.09& \\
IC 5353        &3&$\cap$& 198.4&0.33&16.77&1.2&1.44& 0.03& 0.17& \\
A0119-M1          &3&$\cap$& 690.0&0.80&18.56&3.1&1.06& 0.06& 0.06& \\
A0168-M1          &3&$\cap$&  51.7&0.06&16.62&0.9&1.02&$-$0.48& 0.19& \\
A0189-M1         &3&$\backslash$&  32.3&0.05&17.56&9.8&1.22& 0.85& 0.85& \\
A0261-M1         &3&$\backslash$&  44.0&0.05&18.22&10.0&1.47& 0.76& 0.76& \\
A0295-M1         &3&$\cap$& 513.4&0.63&17.93&3.8&1.24& 0.13& 0.13& \\
A0376-M1         &3&$\cap$& 643.5&0.68&18.36&2.4&1.28& 0.19& 0.20& \\
A0397-M1         &3&$\cap$& 683.0&1.06&18.02&2.7&1.50& 0.07& 0.07& \\
A0419-M1         &3&$\backslash$&  37.8&0.10&17.77&0.6&1.64& 0.33& 0.60& \\
A0496-M1         &3&$\cap$& 449.1&0.73&18.20&2.2&1.01& 0.10& 0.10& \\
A0533-M1         &3&$\cap$& 269.6&0.31&17.30&1.8&1.31& 0.06& 0.10& \\
A0548-M1       &3&$\cap$& 295.9&0.40&17.50&1.0&1.38& 0.00& 0.16& \\
A0634-M1         &3&$\cap$& 130.8&0.23&17.10&3.0&0.88&$-$0.05&$-$0.04& \\
A0912-M1         &3&$\backslash$&  42.7&0.08&16.78&1.7&1.34& 0.48& 0.55& \\
A0999-M1         &3&$\cap$& 387.5&0.62&17.50&1.1&1.65&$-$0.06& 0.05& \\
A1020-M1         &4&$\cap$& 283.6&0.21&17.06&1.7&1.47&$-$0.09& 0.25& \\
A1631-M1         &3&$\cap$& 171.3&0.20&16.65&1.5&1.29&$-$0.03& 0.12& \\
A1831-M1         &4&$\cap$& 762.6&0.49&18.63&5.3&1.11& 0.12& 0.12& \\
A1983-M1         &3&$\wedge$&  85.4&0.10&15.79&0.3&2.63&-1.37& 0.44& \\
A2040-M1         &3&$\cap$& 333.9&0.39&17.80&1.7&1.39& 0.16& 0.19& \\
A2052-M1         &4&$\cap$& 234.1&0.32&18.44&4.5&0.73&$-$0.08& 0.01& \\
A2052-M1         &4&$\cap$& 234.1&0.32&18.44&4.5&0.73&$-$0.08& 0.01& \\
A2147-M1        &3&$\cap$&1500.0&2.21&19.47&1.6&1.37& 0.18& 0.18& \\
A2247-M1        &3&$\backslash$&  39.6&0.03&19.39&1.9&1.42& 0.85& 0.85& \\
A3144-M1         &3&$\cap$& 327.1&0.39&17.19&1.8&1.64& 0.07& 0.11& \\
A3376-M1       &3&$\cap$&1663.9&1.95&19.10&3.1&1.50& 0.05& 0.05& \\
A3395-M1       &3&$\cap$& 333.8&0.37&18.09&2.4&0.98& 0.03& 0.04& \\
A3528-M1       &3&$\cap$& 605.1&0.61&17.81&2.5&1.35& 0.18& 0.18& \\
A3532-M1         &3&$\cap$& 434.9&0.43&17.68&3.1&1.30& 0.18& 0.18& \\
A3554-M1       &3&$\cap$& 456.1&0.52&18.42&2.7&1.10& 0.04& 0.04& \\
A3556-M1       &3&$\cap$& 417.0&0.47&17.16&2.2&1.34& 0.09& 0.10& \\
A3558-M1       &3&$\cap$&1516.4&1.73&19.37&2.1&1.10& 0.05& 0.05& \\
A3562-M1       &3&$\cap$&1034.8&1.15&19.13&1.2&1.32& 0.00& 0.03& \\
A3564-M1         &3&$\cap$& 216.7&0.24&16.96&1.3&1.38& 0.05& 0.20& \\
A3570-M1       &3&$\cap$& 179.4&0.26&16.56&1.3&1.50& 0.00& 0.17& \\
A3571-M1       &3&$\cap$& 843.5&1.15&19.28&2.9&0.75& 0.02& 0.02& \\
A3677-M1         &3&$\cap$& 244.3&0.27&16.89&1.3&1.63&$-$0.04& 0.13& \\
A3716-M1       &3&$\cap$& 384.4&0.44&18.02&2.4&1.10&$-$0.03&$-$0.02& \\
A3736-M1       &3&$\cap$& 875.1&0.92&18.29&1.4&1.33& 0.11& 0.13& \\
A3747-M1       &3&$\cap$& 112.2&0.18&16.04&2.1&1.16&$-$0.02& 0.05& \\
ESO 378-20     &2&$\backslash$&  11.6&2.41&18.22&0.4&2.00& 0.86& 1.09& \\
ESO 443-39     &2&$\backslash$&  11.4&1.30&17.93&1.4&1.31& 0.75& 0.77& \\
ESO 447-30     &2&$\backslash$&  10.1&1.50&17.34&2.7&1.72& 0.84& 0.84& \\
ESO 462-15    &1&$\backslash$&  15.6&0.41&16.59&0.2&1.77&$-$0.10& 0.56& \\
ESO 507-27     &2&$\backslash$&  12.1&3.89&18.57&0.6&1.58& 0.70& 0.79& \\
ESO 507-45   &5&$\wedge$& 117.3&0.33&15.34&1.3&1.26& 0.16& 0.35&$\cap$ in ref. (5)\\
MCG 11-14-25A   &2&$\cap$& 133.0&0.56&17.20&0.7&2.13& 0.00& 0.30&$\wedge$ in ref. (2) \\
MCG 08-27-18    &2&$\backslash$&  11.8&1.95&19.22&0.8&1.93& 0.79& 0.89& \\
UGC 04551       &2&$\backslash$&   6.6&2.26&17.67&2.2&2.16& 0.51& 0.51& \\
UGC 04587       &2&$\backslash$&  11.1&0.19&16.09&1.0&1.24& 0.72& 0.81& \\
UGC 06062       &2&$\backslash$&  10.2&2.75&18.96&0.9&1.81& 0.80& 0.82& \\
VCC 1199       &4&$\backslash$&   3.5&1.60&19.72&0.3&2.96&$-$0.08& 0.90& \\
VCC 1440       &4&$\backslash$&   3.5&0.15&16.86&0.2&1.71& 0.14& 0.89& \\
VCC 1545       &4&$\backslash$&   3.5&2.50&20.43&0.3&1.78& 0.05& 0.51& \\
VCC 1627       &4&$\backslash$&   3.5&0.56&17.89&0.4&2.22&$-$0.10& 0.69& \\
\enddata
\label{tab:cen}
\tablecomments{The profile type, P, is $\backslash=$ power-law,
$\wedge=$ intermediate form, and $\cap=$ core.  The source column refers
to the origin of central structural
parameters for the given galaxy as follows: 1) \citet{l05}; 2) \citet{rest};
3) \citet{laine};
4) \citet{l95} or \citet{f97}; 5) \citet{quil}; and 6) \citet{rav}. The
notes column gives previous profile classifications and their source
if different from the present classification.}
\end{deluxetable}

\begin{deluxetable}{lcccccccccc}
\tablecolumns{11}
\tablewidth{0pt}
\tablecaption{Bimodal Distributions of $\gamma'$}
\tablehead{\colhead{Sample}&\colhead{N}
&\colhead{$f_1$}&\colhead{$f_2$}&\colhead{$\bar\gamma_1'$}
&\colhead{$\bar\gamma_2'$}&\colhead{$\sigma_1$}&\colhead{$\sigma_2$}
&\colhead{$P_1(0.95)$}&\colhead{$P_1=P_2$}&\colhead{$P_2(0.95)$}}
\startdata
Full Sample&223&0.576&0.424&0.13&0.73&0.14&0.14&0.23&0.44&0.52 \\
Full Sample&223&0.535&0.465&0.11&0.69&0.09&0.19&0.23&0.31&0.37 \\
$D\leq50$ Mpc&144&0.413&0.587&0.16&0.73&0.16&0.16&0.30&0.43&0.55 \\
$D\leq50$ Mpc&144&0.377&0.623&0.13&0.72&0.11&0.18&0.24&0.33&0.43 \\
$M_V=-21.5\pm1.0$&\ 97&0.585&0.415&0.17&0.73&0.16&0.16&0.31&0.49&0.57 \\
$M_V=-21.5\pm1.0$&\ 97&0.472&0.528&0.12&0.65&0.10&0.24&0.16&0.32&0.43 \\
\enddata
\label{tab:bimodal}
\tablecomments{For each sample listed, the distribution of $\gamma'$ is
parameterized as the sum of two gaussian distributions. The ``1'' subscript
refers to the gaussian describing the nominal core galaxies, while ``2''
corresponds to the power-law galaxies. The fraction of the sample in
each is given by $f_1$ and $f_2,$ the sample means are $\bar\gamma_1'$
and $\bar\gamma_2',$ and the dispersions are $\sigma_1$ and $\sigma_2.$
Note that in the first distribution listed for each sample both dispersions
are constrained to have the same value.  The last three columns give
the $\gamma'$ values at which the probability of being in the core
populations is 95\%, the probabilities of being in either population
are equal, and the probability of being a power-law galaxy is 95\%.}
\end{deluxetable}

\begin{figure}
\plotone{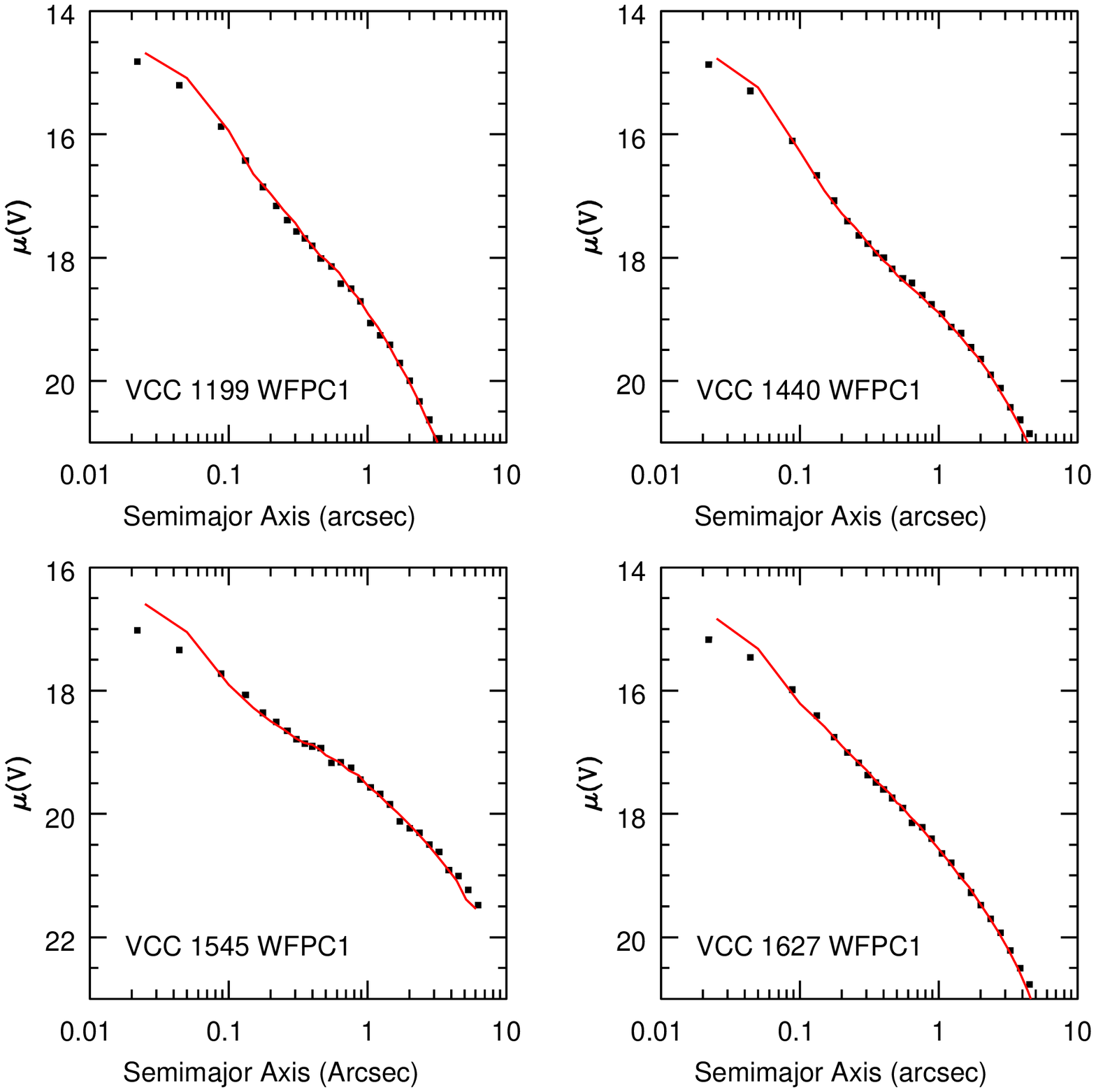}
\end{figure}

\begin{figure}
\plotone{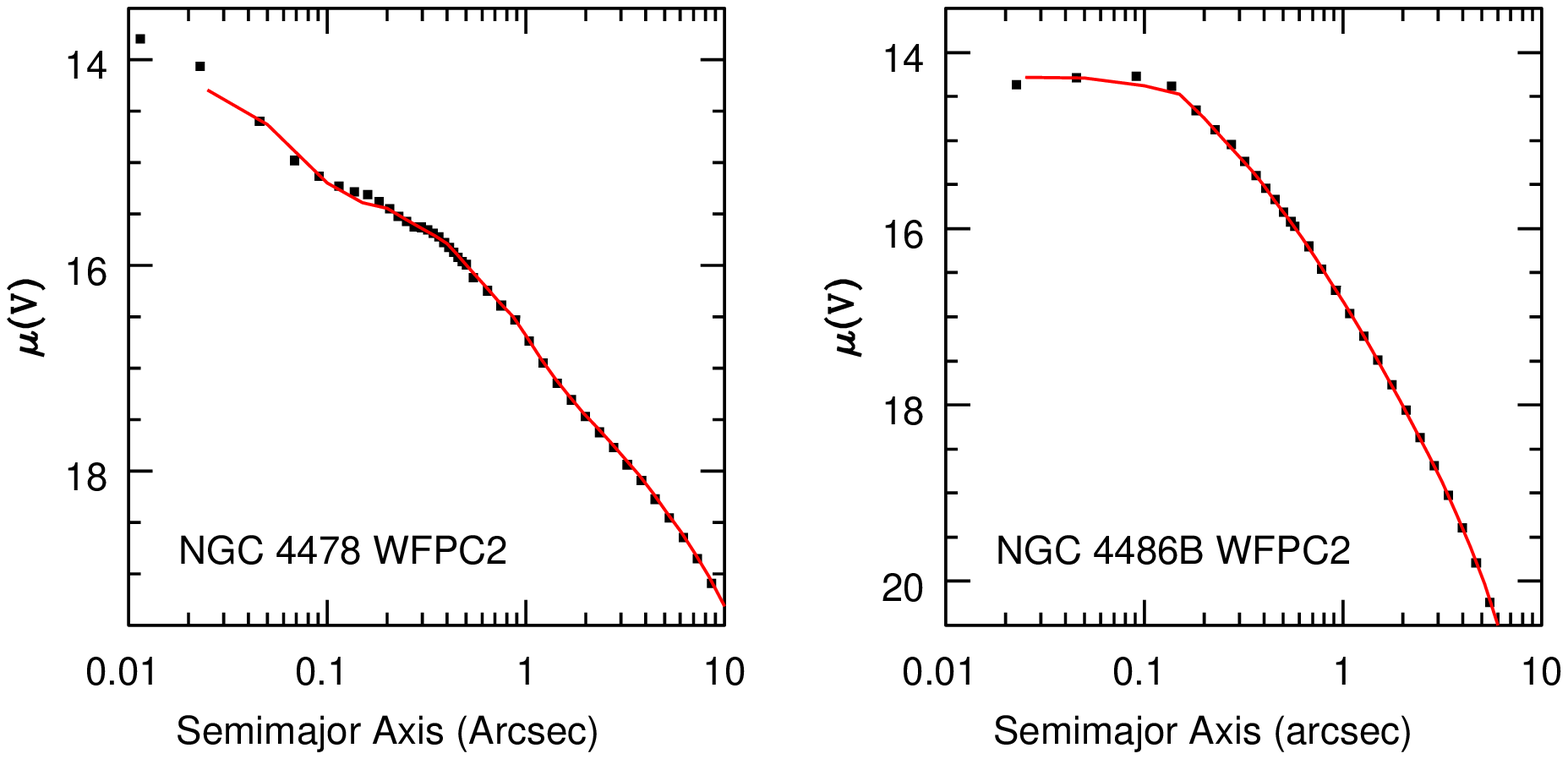}
\end{figure}
\clearpage

\begin{figure}
\includegraphics[scale=0.8]{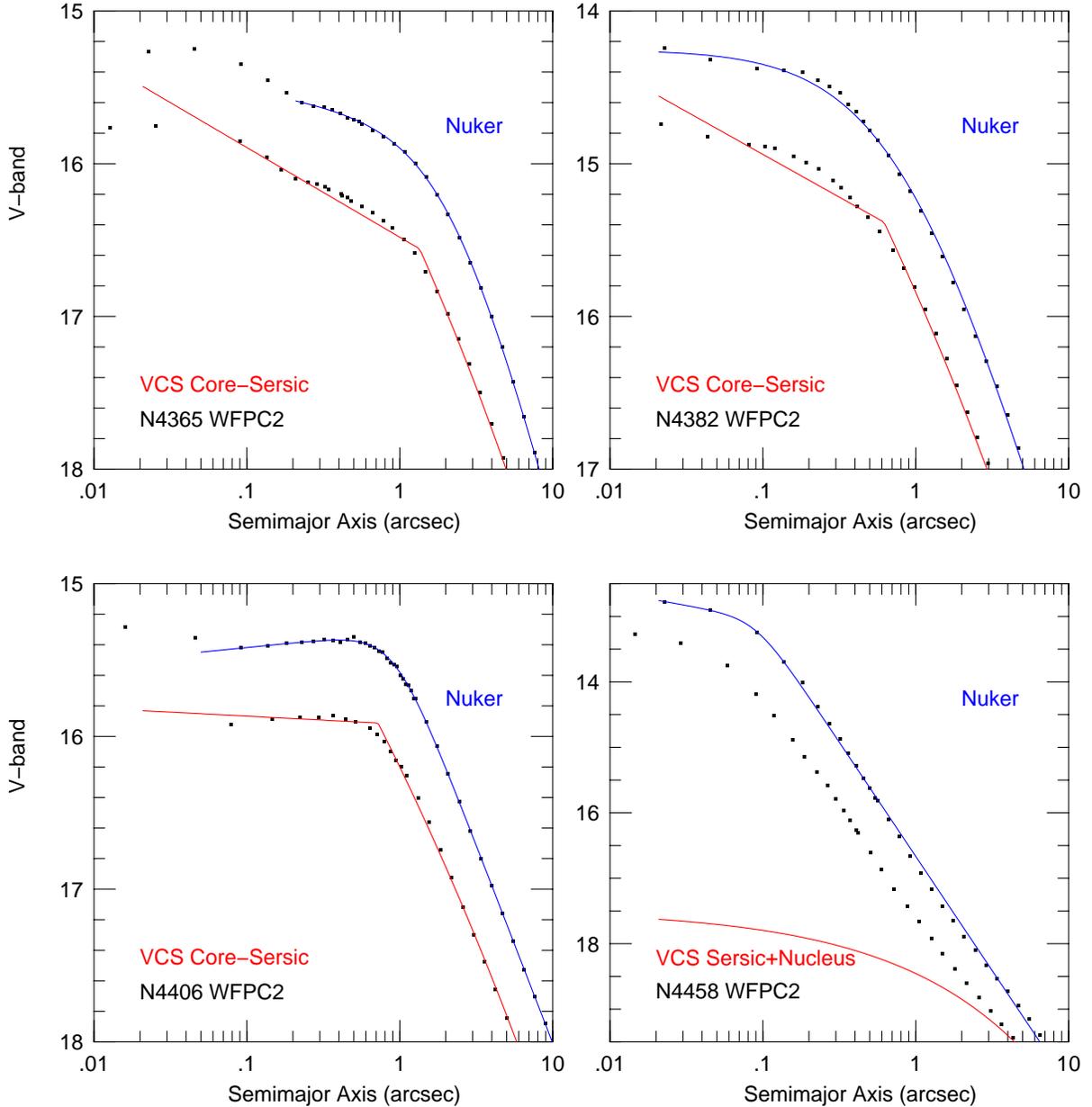}
\caption{Nuker law fits (blue) are compared to the \citet{lf}
model fits (red) for core galaxies in common between
the present and VCS samples.  The points are deconvolved isophotal
surface brightness measurements obtained with WFPC2 \citep{l05},
with the exception of NGC 4486, for which WFPC1 was used \citep{l92a}.
The photometry has been recast as a function of isophote mean-radius
for comparison to the VCS models.  The VCS models have been scaled
vertically to account for the $g-V$ color difference by computing the
mean offset between the data and models for $r>2''.$  The data used
for comparison to the Nuker models has been offset by 0.5 mag for clarity.
The legends in red indicate the type of the VCS model used, which is
either a Core-\Ser model, a pure \Ser model, or a \Ser model
with an additional nuclear component.
The nuclear components in the last model are not shown, however, as
the \citet{lf} estimation of $\gamma'$ is based solely on the \Ser component.}
\label{fig:acscomp1}
\end{figure}

\begin{figure}
\plotone{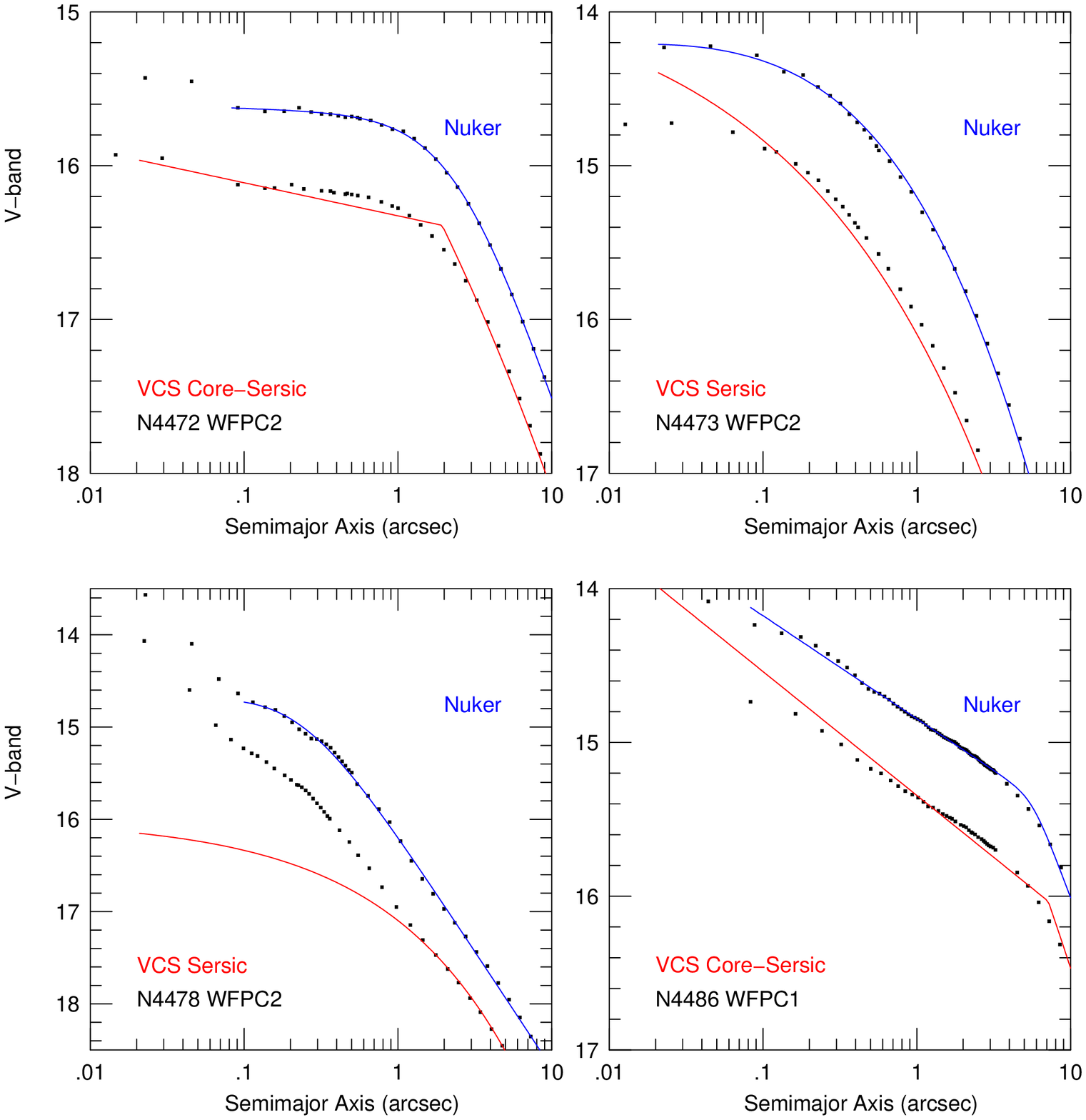}
\end{figure}

\begin{figure}
\plotone{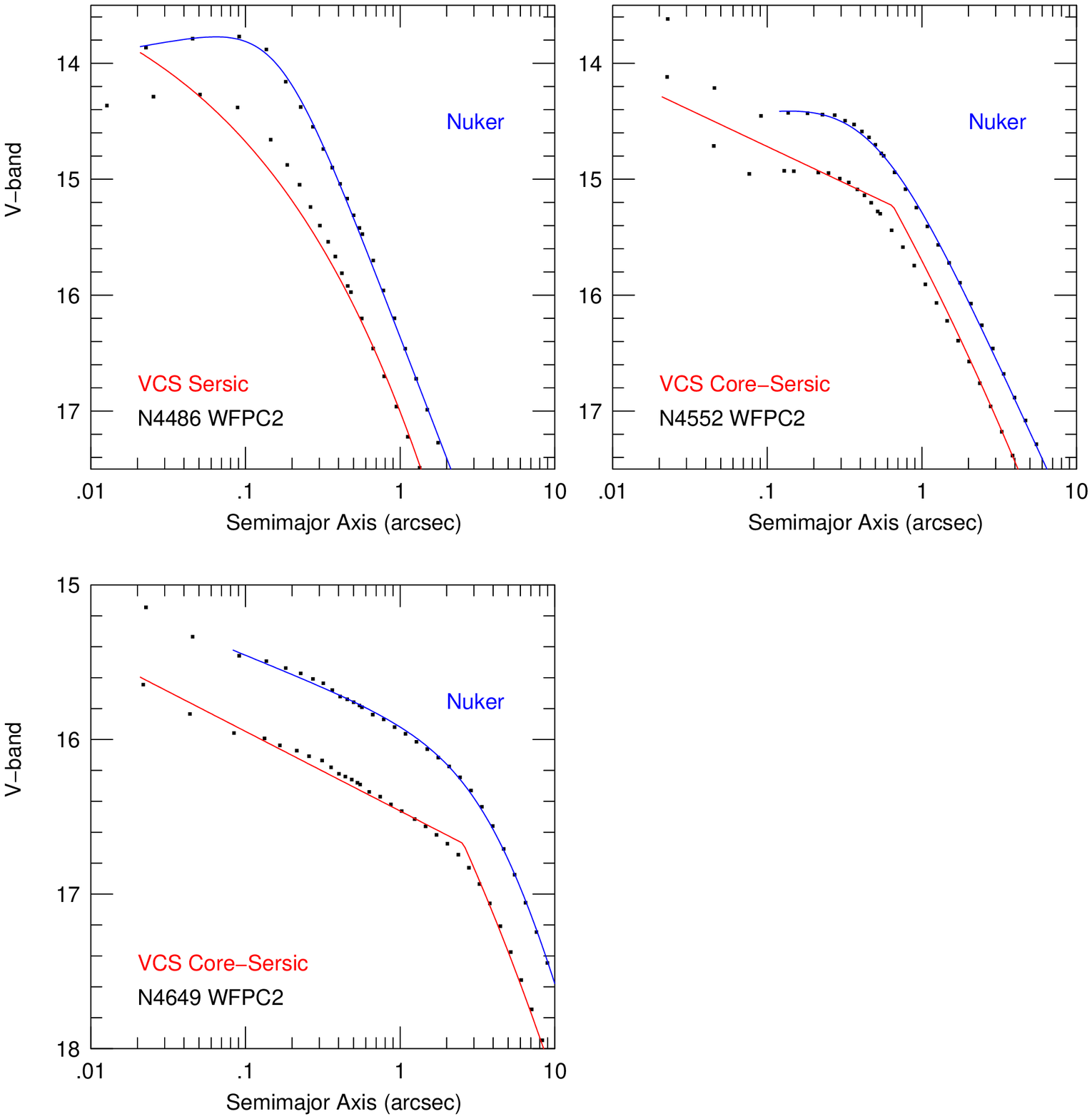}
\end{figure}

\begin{figure}
\plotone{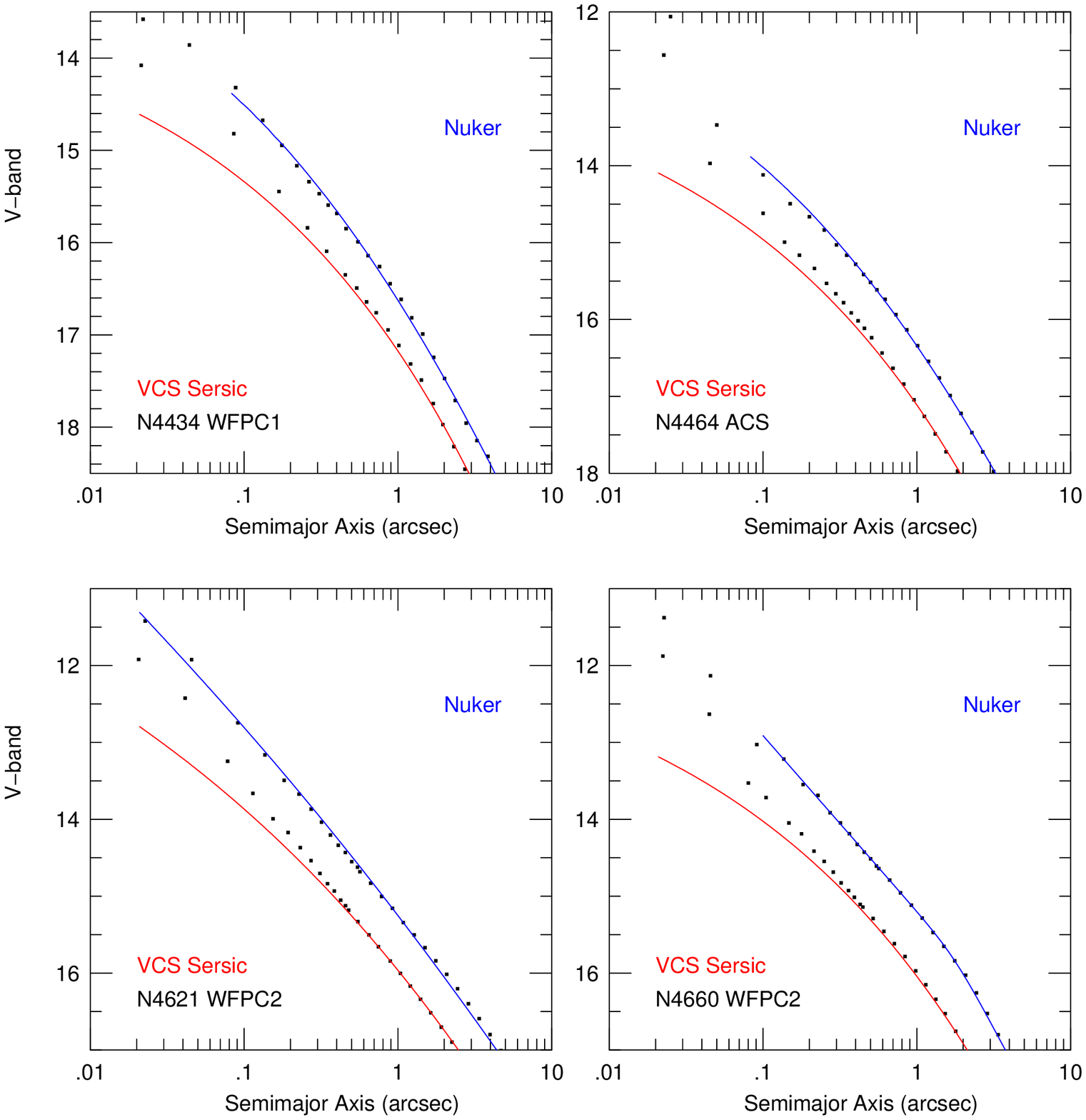}
\caption{Nuker law fits (blue) are compared to the \citet{lf}
model fits (red) for power-law galaxies in common between
the present and VCS samples that \citet{lf} fitted with pure \Ser models.
The points are deconvolved isophotal surface brightness
measurements obtained either with WFPC2 \citep{l05}
for NGC 4621 and 4660, or WFPC1 \citep{l95} for NGC 4434,
and ACS/WFC for NGC 4464.
The photometry has been recast as a function of isophote mean-radius
for comparison to the VCS models.  The VCS models have been scaled
vertically to account for the $g-V$ color difference by computing the
mean offset between the data and models for $r>2''.$  The data used
for comparison to the Nuker models has been offset by 0.5 mag for clarity.}
\label{fig:acscomp2}
\end{figure}

\begin{figure}
\plotone{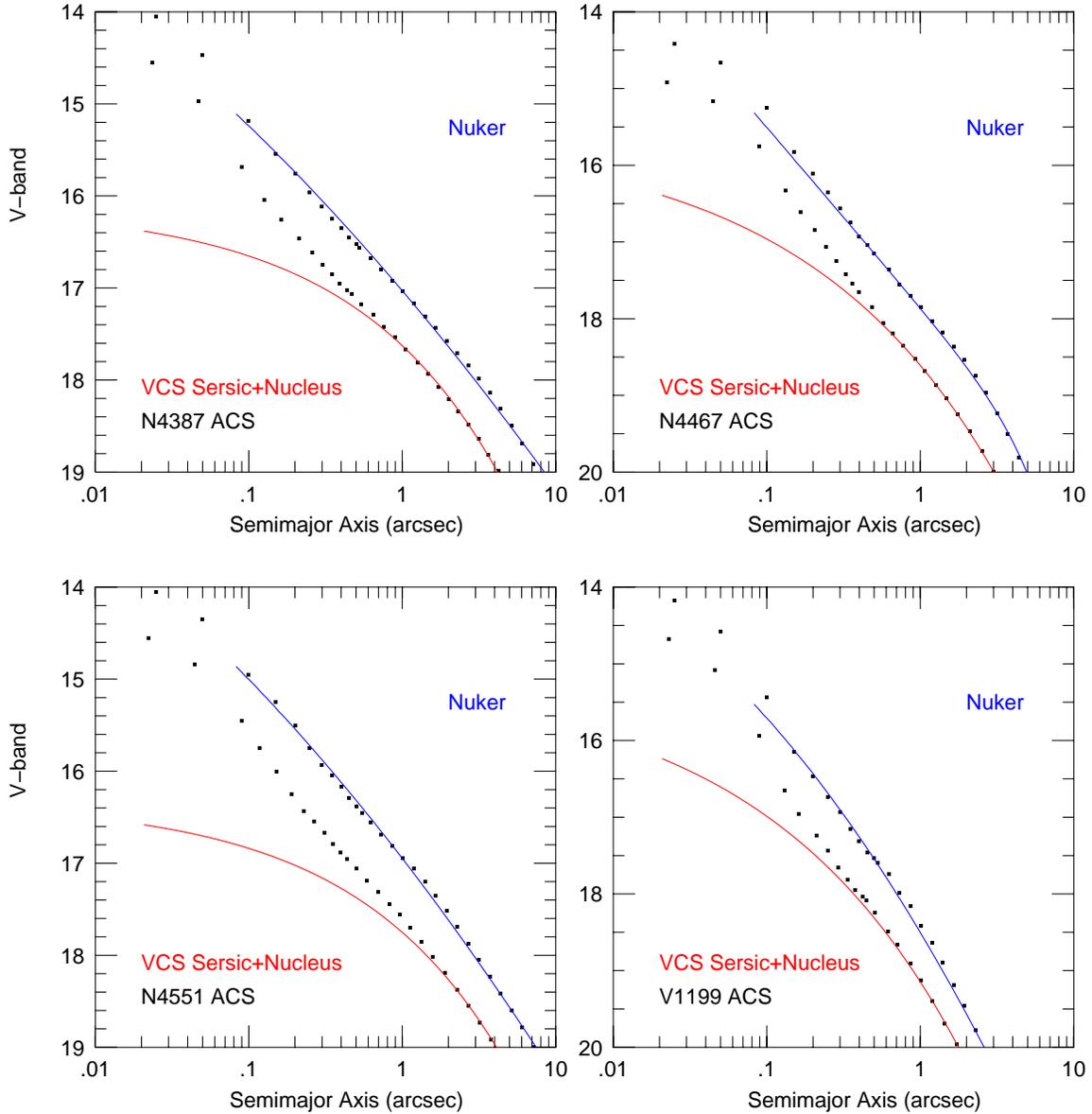}
\caption{Nuker law fits (blue) are compared to the \citet{lf}
model fits (red) for power-law galaxies in common between
the present and VCS samples that \citet{lf} fitted with \Ser $+$
nucleus models.
The points are deconvolved isophotal surface brightness
measurements obtained from ACS/WFC images.
The photometry has been recast as a function of isophote mean-radius
for comparison to the VCS models.  The VCS models have been scaled
vertically to account for the $g-V$ color difference by computing the
mean offset between the data and models for $r>2''.$  The data used
for comparison to the Nuker models has been offset by 0.5 mag for clarity.
The nucleus components fitted by \citet{lf} are not shown as
their estimation of $\gamma'$ is based solely on the models shown.}
\label{fig:acscomp3}
\end{figure}

\begin{figure}
\plotone{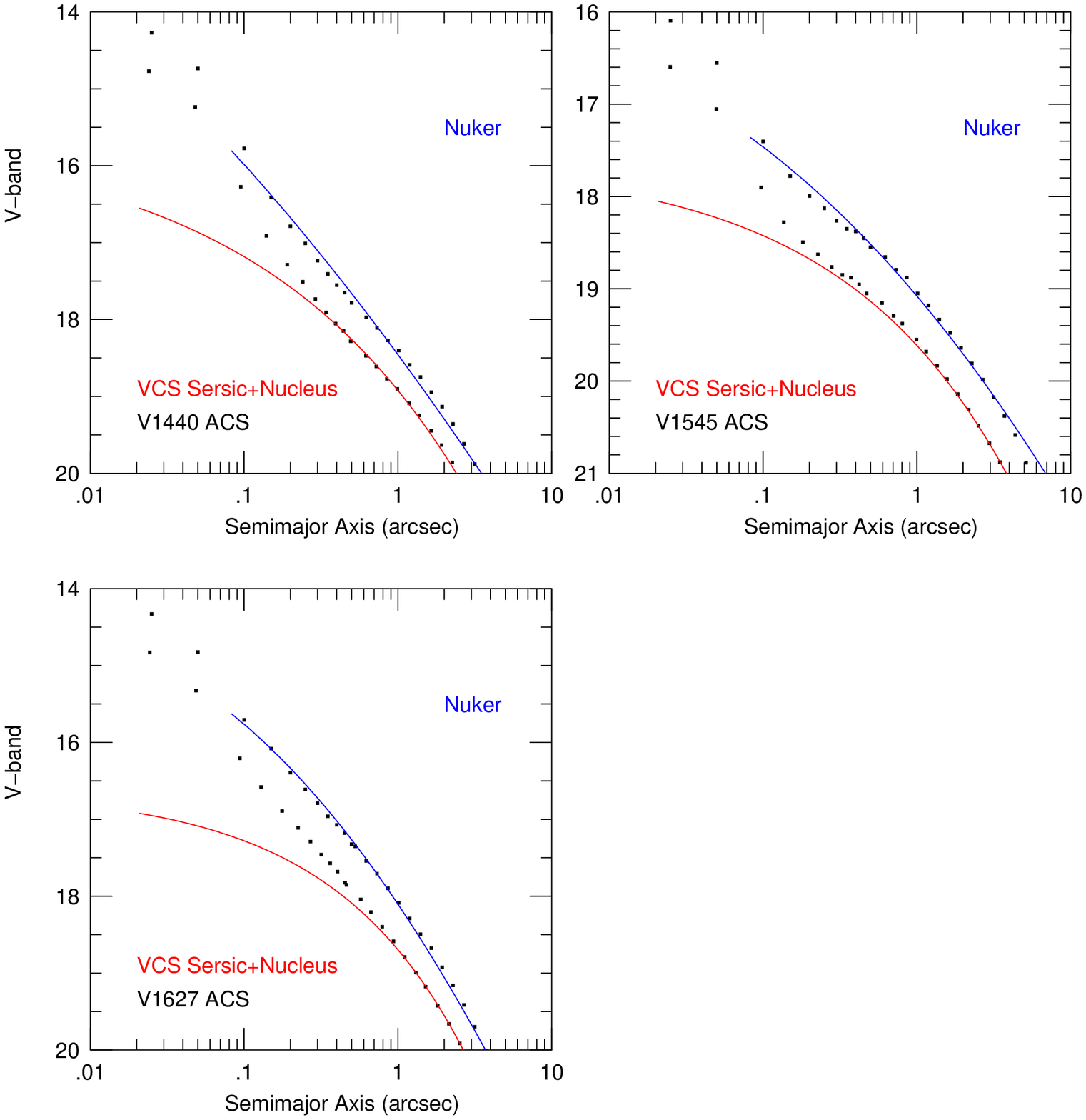}
\end{figure}

\begin{figure}
\plotone{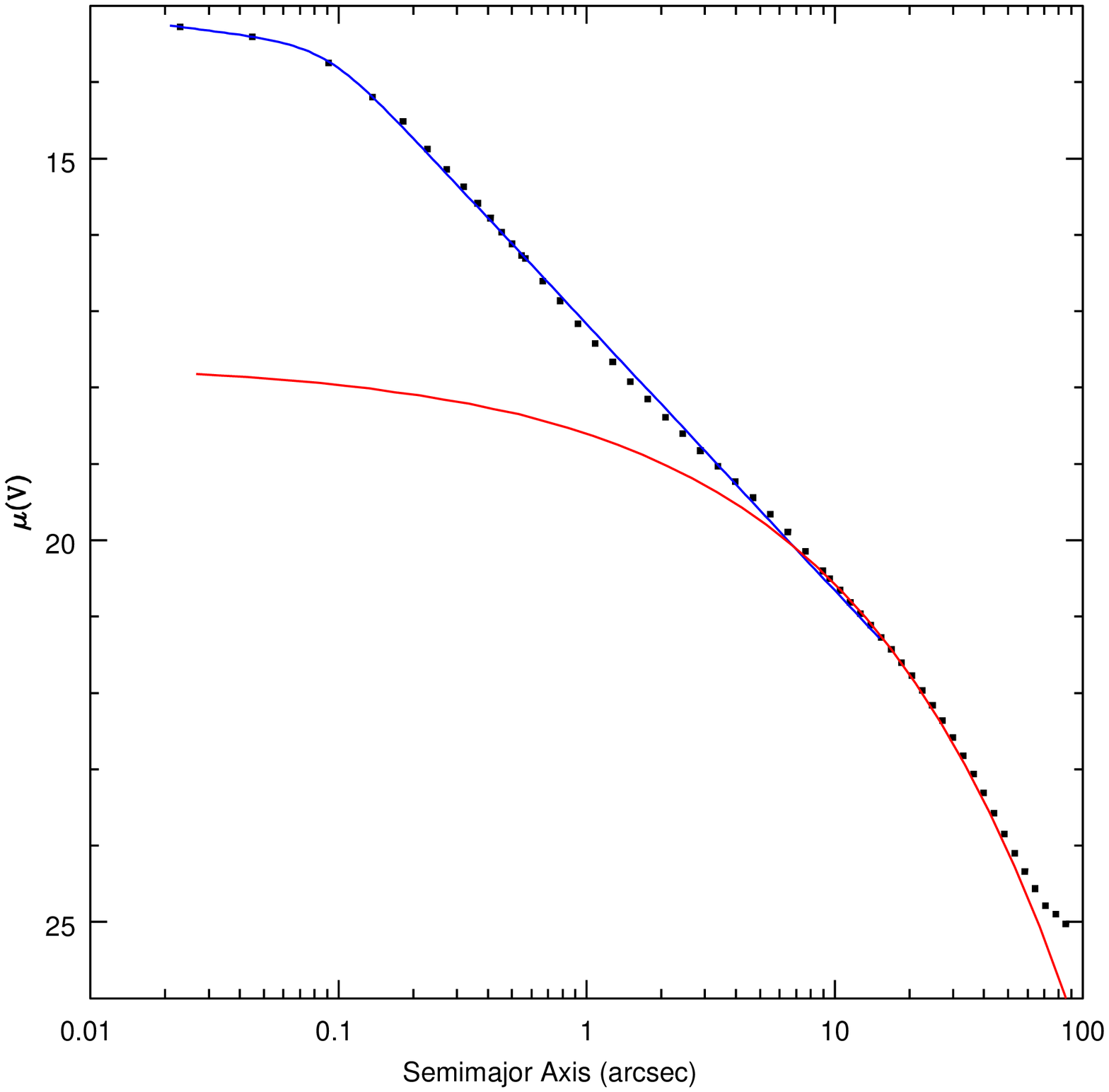}
\caption{The VCS and present modeling procedures are compared for
NGC 4458.  The \citet{l05} deconvolved WPFC2 $V-$band surface photometry of
NGC 4458 extended to large radii by the addition of the \citet{jed} $B-$band
photometry for $r>10''$ is plotted as black points.
The Nuker law is fitted to NGC 4458 for $r<15''$ and
is plotted in blue, while the \citet{lf} \Ser {\it g}-band component
of their \Ser $+$ Nucleus model is plotted in red.
The Nuker law describes the portion of the galaxy classified as a
separate nuclear component in the VCS model, while the \Ser
component alone is used to estimate the VCS $\gamma'$ value at $0\asec1.$
The \Ser profile is adjusted from a geometric
radius system to semimajor axis by the average isophote ellipticity
at $r\sim10$ ($\epsilon=0.12$) and is scaled vertically
assuming $g-V=0.35$ mag.}
\label{fig:n4458}
\end{figure}

\end{document}